\documentclass[aps,prd,twocolumn,superscriptaddress,nofootinbib]{revtex4-1}

\usepackage{amsmath,amssymb,amsfonts}
\usepackage{graphicx}
\usepackage{dcolumn}
\usepackage{bm}
\usepackage{hyperref}
\usepackage[sort&compress]{natbib}
\usepackage{color}

\newcommand{\mbf}{\mathbf}

\begin{document}

\title{Universal self-similar dynamics of relativistic and nonrelativistic field theories near nonthermal fixed points}

\author{Asier Pi\~neiro Orioli}
   \email{pineiroorioli@thphys.uni-heidelberg.de}
\affiliation{Institut f\"ur Theoretische Physik, Universit\"at Heidelberg, Philosophenweg 16, 69120 Heidelberg, Germany}
\author{Kirill Boguslavski}
   \email{K.Boguslavski@thphys.uni-heidelberg.de}
\author{J\"urgen Berges}
   \email{J.Berges@thphys.uni-heidelberg.de}
\affiliation{Institut f\"ur Theoretische Physik, Universit\"at Heidelberg, Philosophenweg 16, 69120 Heidelberg, Germany}
\affiliation{ExtreMe Matter Institute EMMI, Planckstra\ss e 1, 64291 Darmstadt, Germany}

\begin{abstract}
We investigate universal behavior of isolated many-body systems far from equilibrium, which is relevant for a wide range of applications from ultracold quantum gases to high-energy particle physics. The universality is based on the existence of nonthermal fixed points, which represent nonequilibrium attractor solutions with self-similar scaling behavior. The corresponding dynamic universality classes turn out to be remarkably large, encompassing both relativistic as well as nonrelativistic quantum and classical systems. For the examples of nonrelativistic (Gross-Pitaevskii) and relativistic scalar field theory with quartic self-interactions, we demonstrate that infrared scaling exponents as well as scaling functions agree. We perform two independent nonperturbative calculations, first by using classical-statistical lattice simulation techniques and second by applying a vertex-resummed kinetic theory. The latter extends
kinetic descriptions to the nonperturbative regime of overoccupied modes.
Our results open new perspectives to learn from experiments with cold atoms aspects about
the dynamics during the early stages of our universe. 
\end{abstract}

\maketitle

\section{Introduction}
\label{sec:intro}

\subsection{Universality far from equilibrium}

The notion of universality or scaling phenomena
far from equilibrium is to a large extent unexplored, in particular, in
relativistic quantum field theories. Here the strong interest is mainly driven by advances in our understanding of the early universe dynamics as well as relativistic collision experiments of heavy nuclei in the laboratory.
Experimentally, the investigation of quantum systems far from equilibrium is also boosted by the nonrelativistic physics of ultracold quantum gases. By using optical or atom chip traps, they provide a flexible testbed with tunable interactions, symmetries and dimensionality, with connections to a wide variety of physical systems. This may include high-energy physics, in particular, if both relativistic and nonrelativistic systems belong to the same universality class~\cite{Berges:2014bba}. It is the aim of this paper to establish such a connection for bosonic field theories far from equilibrium.

In recent years, important new universality classes have been discovered in isolated relativistic systems far from equilibrium~\cite{Micha:2002ey,Micha:2004bv,Berges:2008wm,Berges:2008mr,Berges:2010ez,Berges:2010zv,Gasenzer:2011by,Schlichting:2012es,Kurkela:2012hp,Berges:2013eia,Gasenzer:2013era,Berges:2013fga,Berges:2013lsa,Berges:2013oba,York:2014wja,Kurkela:2014tea}. The universality is based on the existence of nonthermal fixed points~\cite{Berges:2008wm,Berges:2008sr,Scheppach:2009wu}, which represent nonequilibrium attractor solutions with self-similar scaling behavior. The underlying physical processes are reminiscent of the stationary transport of conserved charges in the phenomenon of wave turbulence~\cite{Zakharov:1992,Nazarenko:2011}. However, no external sources or sinks are present for isolated systems. In high-energy physics, collision experiments of heavy nuclei (where a `fireball' expands against the surrounding vacuum) and the evolution of the early universe provide important examples for isolated quantum systems.  

\begin{figure}[t!]
\centering
	\includegraphics[width=.5\textwidth]{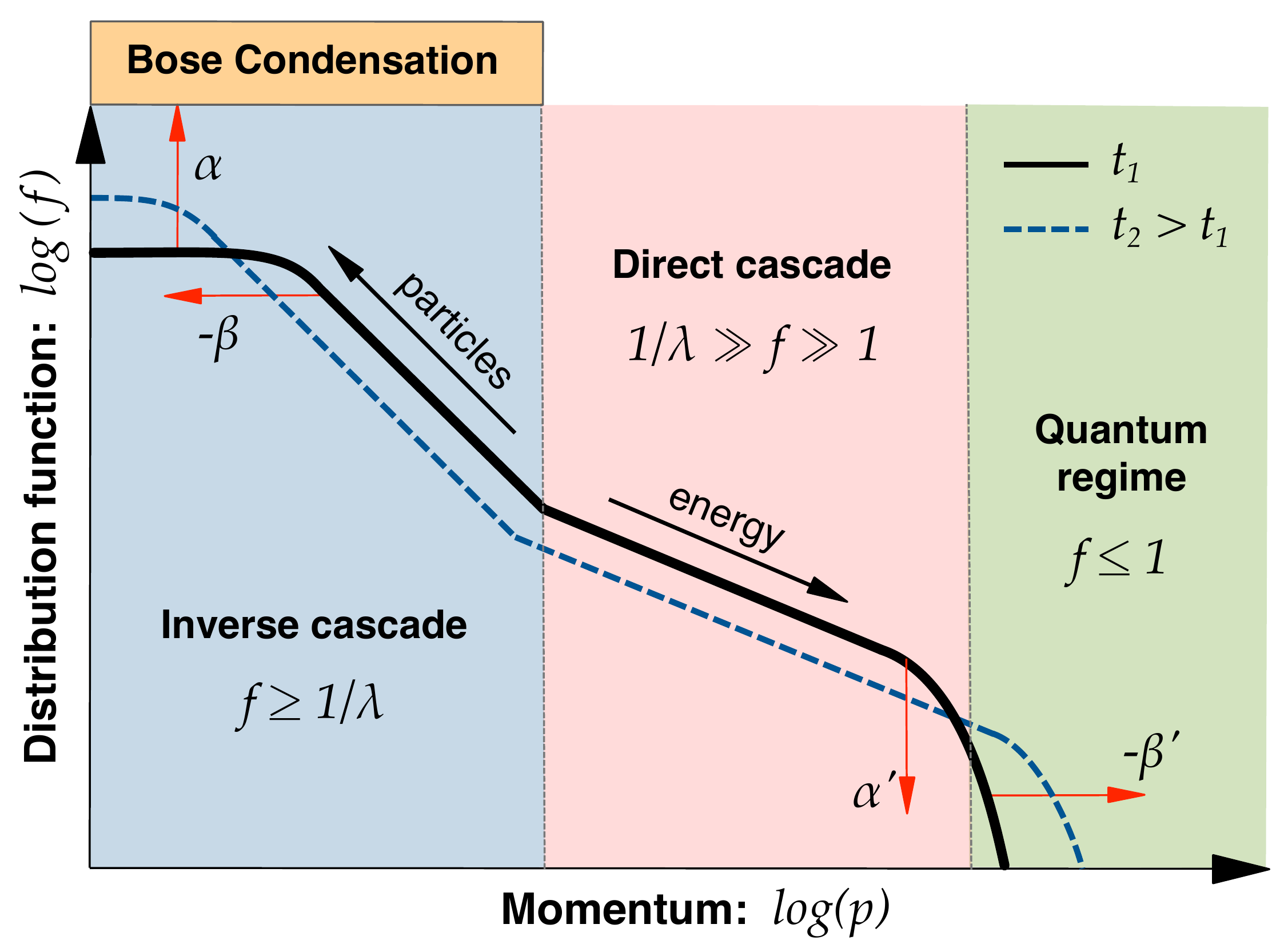}
	\caption{Schematic illustration of the occupation number distribution near a nonthermal fixed point, shown as a function of momentum $p$ for two subsequent times $t_1$ and $t_2 > t_1$. The scaling exponents $\alpha$ and $\beta$ characterize the self-similar evolution according to equation (\ref{eq:selfsim}).}
	\label{fig:dual_cascade}
\end{figure}

Also experimental setups employing ultracold quantum gases can be largely isolated, such that their dynamics is governed by unitary time evolution. Nonthermal fixed points have been investigated in this context for Bose gases~\cite{Scheppach:2009wu,Nowak:2010tm,Nowak:2011sk,Berges:2012us,Schole:2012kt,Nowak:2012gd,Karl:2013kua,Berges:2014xea,Ewerz:2014tua}. In these nonrelativistic systems, Bose condensation occurs out-of-equilibrium as a consequence of an inverse particle cascade~\cite{Svistunov:1,Kagan:1,Berloff:1,Berges:2012us,Nowak:2012gd}. Similar cascades with nonthermal Bose condensation are also known for relativistic scalar field theories in various spatial dimensions~\cite{Berges:2008wm,Berges:2008sr,Berges:2010ez,Gasenzer:2011by,Berges:2012us}. The question arises, whether these similarities between the relativistic and nonrelativistic systems can be made quantitative. Clearly, there are important differences and one has to specify which properties can be universal.  

To illustrate this, FIG.~\ref{fig:dual_cascade} shows a typical distribution function $f(t,{\mbf p})$ near a nonthermal fixed point as a function of momentum ${\mbf p}$ for two subsequent times $t=t_1$ and $t_2 > t_1$. We distinguish different momentum ranges by their occupancies $f(t,{\mbf p})$ in terms of a small (interaction or `diluteness') parameter $\lambda \ll 1$ as will be explained below. The inverse particle cascade leading to Bose condensation occurs in the highly occupied low-momentum regime, where $f(t,{\mbf p}) \gtrsim 1/\lambda$~\cite{Berges:2008wm}. This particle transport towards low momenta is part of a dual cascade, in which energy is also transferred by weak wave-turbulence towards higher momenta ($1/\lambda \gg f(t,{\mbf p}) \gg 1$). The latter evolves until a high-momentum scale is reached, where the characteristic mode occupancy becomes comparable to the `quantum-half'~\cite{Micha:2002ey}. 

The different cascades exhibit approximate power-law behavior for mode occupancies $f(t,{\mbf p})$ in characteristic inertial ranges of momenta. However, in general isolated systems out of equilibrium cannot realize stationary transport solutions in the absence of external driving forces. Instead, the transport in isolated systems is described in terms of the more general notion of a {\it self-similar} evolution, where the distribution function obeys for isotropic systems
\begin{equation}
	f(t,{\mbf p})=t^\alpha\,f_S(\xi \equiv t^\beta |{\mbf p}|) \,
\label{eq:selfsim}
\end{equation}
in a given scaling regime. Here, all quantities are considered to be dimensionless by use of some suitable momentum scale, which is specified below. 

The values of the scaling exponents $\alpha$ and $\beta$, as well as the form of the nonthermal fixed point distribution $f_S(\xi)$ are universal. More precisely, all models in the same universality class can be related by a multiplicative rescaling of $t$ and ${\mbf p}$. Quantities which are invariant under this rescaling are universal. Accordingly, all system-dependent aspects of the distribution are contained in two nonuniversal amplitudes, which we denote as $A$ 
and $B$. It is convenient to define them according to $f_S(\xi=B) = A$ with $d f_S(\xi=B)/d\xi = - 2 A/B$ such that $A$ characterizes the amplitude of the scaling function at $\xi=B$, where the occupation number receives its dominant contribution.   

We emphasize that the universal properties can be different for different inertial ranges. This is indicated in FIG.~\ref{fig:dual_cascade}, where in the direct cascade regime other scaling exponents $\alpha^\prime$, $\beta^\prime$ and a different scaling function $f_S^\prime$ than in the infrared are found. For instance, two theories can have the same universal low-momentum properties while they may differ significantly in another inertial range at higher scales. This is very similar to the classification of universal properties in thermal equilibrium, where one distinguishes for a given theory between infrared and ultraviolet fixed points and associated scaling properties depending on the momentum regime.

\subsection{Outline of results}
\label{sec:outline}

In this work we compute the exponents $\alpha$, $\beta$ and the scaling function $f_S$ of the self-similar distribution (\ref{eq:selfsim}) in the infrared regime. We present results for nonrelativistic (Gross-Pitaevskii) as well as (massless) relativistic scalar field theory with quartic self-interactions, respectively. While the relativistic theory captures important aspects of the Higgs sector of particle physics or of inflationary models for early universe dynamics, the Gross-Pitaevskii field theory can describe a dilute Bose gas. 

A central conclusion of this paper is that the infrared scaling exponents and scaling functions of these theories belong to the same universality class, i.e.~the results for $\alpha$, $\beta$ and the universal form of the scaling function $f_S$ in~(\ref{eq:selfsim}) agree. This is nontrivial, in particular, since the nonrelativistic system conserves total particle number whereas in the relativistic theory number-changing processes are possible. The agreement found in the infrared is also remarkable in view of the fact that in the inertial range of the direct energy cascade towards higher momenta the exponents $\alpha^\prime$, $\beta^\prime$ and the scaling function $f_S^\prime$ from the different theories do not agree~\cite{Micha:2002ey,Micha:2004bv}.  

Since the large occupancies at low momenta lead to strongly nonlinear dynamics, one cannot apply standard perturbative kinetic theory in the infrared. (Perturbative approaches~\cite{Zakharov:1992,Micha:2002ey,Micha:2004bv} are often used to describe the direct energy cascade at higher momenta.\footnote{Perturbative kinetic theory has also been employed in the infrared to describe Bose condensation at low momenta~\cite{Semikoz:1994zp,Semikoz:1995rd}. However, it is known to neglect important vertex corrections in this case~\cite{Berges:2008wm,Berges:2012us} as is explained in section~\ref{sec:vrkt}.}) We perform two independent nonperturbative calculations. The first approach employs classical-statistical lattice simulation techniques in sections \ref{sec:nonrel} and~\ref{sec:rel}~\cite{Khlebnikov:1996mc,Berges:2010ez}. The second, analytical method applies a vertex-resummed kinetic theory in sections~\ref{sec:vrkt} and \ref{sec:anomalousscaling}, which is based on an expansion in the number of field components $N$ to next-to-leading order~\cite{Berges:2001fi,Aarts:2002dj}. The approach extends well-established kinetic descriptions~\cite{Zakharov:1992,Nazarenko:2011,Micha:2004bv} to the nonperturbative regime of overoccupied modes. 

This is used to obtain the analytic estimate for the scaling exponents of the  
\begin{align}
\text{\it particle transport:}\quad	\alpha=&\, \beta\,d\, , \quad \beta=\,\frac{1}{2} \,,
\label{eq:exponentsIPC}
\end{align}
towards low momenta, with spatial dimension $d$. 
This is a central analytic result of this work. In contrast to the previously known negative values for $\alpha$ and $\beta$ from perturbative estimates~\cite{Micha:2002ey,Micha:2004bv}, the positive values obtained from the vertex-resummed kinetic theory describe an {\it inverse} particle transport with growing occupation number in the infrared. The fixed relation between $\alpha$ and $\beta$ reflects the conservation of particle number density $n=\int\mathrm{d}^dp/(2\pi)^d\,f(t,{\mbf p}) \sim t^{\alpha - d \beta}$ in this (nonrelativistic) inertial range by using the self-similarity (\ref{eq:selfsim}). Most notably, the emergence of an effectively conserved particle number plays a crucial role for the nonequilibrium evolution of the relativistic theory. 

\begin{figure}[t!]
\centering
	\includegraphics[width=.5\textwidth]{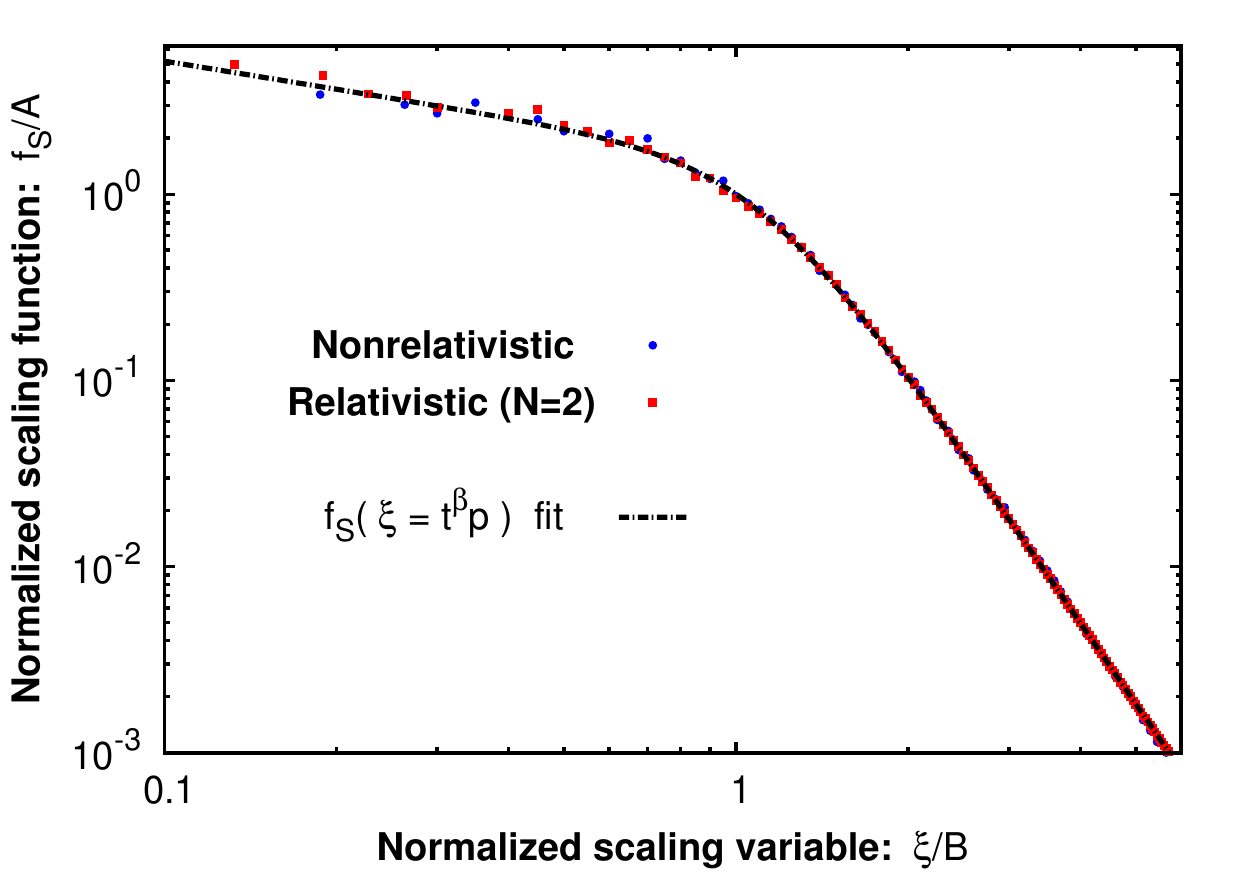}
	\caption{The scaling function $f_S(\xi \equiv t^\beta |{\mbf p}|) = t^{-\alpha} f(t,{\mbf p})$, normalized to the amplitudes $A$ and $B$, exhibits accurate agreement between the nonrelativistic (circles) and relativistic (squares) simulation results.}
	\label{fig:compare_rel_nonrel}
\end{figure}

The quantitative agreement of the NLO estimates of $\alpha = 3/2$ and $\beta = 1/2$ for $d=3$ with the full simulation results for the relativistic (section~\ref{sec:simreldyn}) and for the nonrelativistic (section~\ref{sec:simcl}) theory is remarkable. Though we extend the analytic estimates to include a possible anomalous scaling in section~\ref{sec:anomalousscaling}, we find no strong indications for a nonvanishing anomalous dimension. Furthermore, we analyze the phenomenon of nonthermal Bose condensation from the inverse particle transport towards the zero mode. Most importantly, the dynamical generation of a mass gap $m$ for the (massless) relativistic theory from an intriguing interplay of condensation and medium effects explains why the relativistic theory exhibits nonrelativistic physics at low momenta $|{\mbf p}| \lesssim m$, as is shown in section~\ref{sec:rel}. 

The numerically computed fixed point distribution $f_S$ of equation~(\ref{eq:selfsim}) is given in FIG.~\ref{fig:compare_rel_nonrel}. Shown is the normalized distribution $f_S/A$ as a function of the rescaled variable $\xi/B$, using the model specific amplitudes $A$ and $B$ as defined above. For the relativistic theory, we show results for $N=2$ field components in FIG.~\ref{fig:compare_rel_nonrel} but we consider also other values of $N$ in section~\ref{sec:rel}.
The lattice data for both nonrelativistic and relativistic theories turns out to be well described by the fit (dashed line) 
\begin{equation}
f_S(\xi) \simeq \frac{A (\kappa_> - \kappa_<)}{(\kappa_>-2)(\xi/B)^{\kappa_<} + (2-\kappa_<)(\xi/B)^{\kappa_>}}
\label{eq:fit}
\end{equation}
for $\xi>0$ with approximately $\kappa_< \simeq 0.5$ and $\kappa_> \simeq 4.5$ for the available range of momenta and times. We note that the value for $\kappa_>$ is rather close to those obtained for stationary turbulence in nonrelativistic systems~\cite{Scheppach:2009wu} as shown in appendix~\ref{eq:waveturbulence}. 

The very good agreement of the exponents and nonthermal fixed point distributions for the different theories in the infrared scaling regime is a striking manifestation of universality far from equilibrium. One may use this universality, for instance, to learn from experiments with ultracold atoms aspects about the dynamics during the early stages of our universe~\cite{Schmiedmayer:2013xsa}.

\section{Nonrelativistic Bose gas\label{sec:nonrel}}

\subsection{Initial conditions}

We first consider a nonrelativistic atomic gas of interacting bosons with s-wave scattering length $a$.\footnote{Here, three spatial dimensions are considered and natural units will be employed where the reduced Planck constant ($\hbar$), the speed of light ($c$) and Boltzmann's constant ($k_B$) are set to one.} For a gas of density $n$ the average interatomic distance is $n^{-1/3}$. Together with the scattering length $a$, this can be used to define a dimensionless `diluteness parameter'
\begin{equation}
\zeta = \sqrt{n a^3} \, .
\label{eq:dilute}
\end{equation}
For a typical scattering length of, e.g., $a \simeq 5 \,{\rm nm}$ and bulk density $n \simeq 10^{14} \,{\rm cm}^{-3}$ the diluteness parameter $\zeta \simeq 10^{-3}$ is very small, and in the following we will always assume $\zeta \ll 1$. The density and scattering length can also be used to define a characteristic `coherence length', whose inverse is described by the momentum scale 
\begin{equation}
Q = \sqrt{16\pi a n} \, .
\label{eq:Q}
\end{equation}

To observe the dynamics near nonthermal fixed points for the interacting Bose gas, an unusually large occupancy of modes at the inverse coherence length scale $Q$ has to be prepared. Such nonequilibrium extreme conditions may be obtained, for instance, from a quench or nonequilibrium instabilities~\cite{Scheppach:2009wu,Nowak:2010tm,Nowak:2011sk,Berges:2012us,Schole:2012kt,Nowak:2012gd,Karl:2013kua,Berges:2014xea}. More precisely, for a weakly coupled gas of average density $n = \int d^3p/(2\pi)^3 f(|{\mbf p}|)$ this requires a large characteristic mode occupancy 
\begin{equation}
f(Q) \sim \frac{1}{\zeta} \gg 1\, . 
\label{eq:overpopulation}
\end{equation}
This represents an extreme nonequilibrium distribution of modes. The large typical occupancies lead to essentially classical dynamics. Most importantly, the system in this overoccupied regime is strongly correlated. 

These properties may be understood from a Gross-Pitaevskii equation for a nonrelativistic complex Bose field $\psi$: 
\begin{align}
	i\partial_t\psi(t,\mbf x)=&\, \left( -\frac{\nabla^2}{2m} + g|\psi(t,\mbf x)|^2 \right)\psi(t,\mbf x).
\label{eq:gpe}
\end{align}
The coupling $g$ is not dimensionless and determined from the mass $m$ and scattering length as $g = 4\pi a/m$. The total number of particles, given by $N_{\rm total} = \int d^3 x |\psi(t, \mbf x)|^2$, is conserved. 

In the mean-field approximation the effect of the interaction term in the Gross-Pitaevskii
equation is a constant energy shift for each particle,
\begin{equation}
\Delta E = 2 g \langle |\psi|^2 \rangle = 2 g n = 2 g \int \frac{d^3p}{(2\pi)^3} f(|{\mbf p}|) \, , 
\end{equation}
which can be absorbed in a redefinition of the chemical potential. However, we note that for the very high occupancy (\ref{eq:overpopulation}) of the typical momentum $Q$ the shift in energy is not small compared to the relevant kinetic energy, i.e.\ $2 g n \sim Q^2/2m$.  Parametrically, this can be directly verified using (\ref{eq:overpopulation}):
\begin{equation}
g \int d^3p\, f(|{\mbf p}|) \sim g\, Q^3 f(Q) \sim g \frac{Q^3}{\zeta} \sim g \frac{Q^3}{m g Q} \sim \frac{Q^2}{m} \, .
\end{equation}
Here we have used that with $a = m g/(4\pi)$ equation (\ref{eq:Q}) implies $Q = 2 \sqrt{m g n}$ and (\ref{eq:dilute}) gives $\zeta = m g Q/(16 \pi^{3/2})$. Most importantly, the energy shift $2gn$ is of the order of the kinetic energy $Q^2/2m$ irrespective of the coupling strength $g$. This already hints at a strongly correlated system, where the dependence on the details of the underlying model parameters is lost.

\subsection{Self-similar dynamics from classical-statistical simulations}
\label{sec:simcl}

In order to perform simulations beyond the mean-field approximation, we use the fact that the nonequilibrium quantum dynamics of the highly occupied system can be accurately mapped onto a classical-statistical field theory evolution~\cite{Son:1996zs,Khlebnikov:1996mc,Aarts:2001yn,Arrizabalaga:2004iw,Berges:2007ym}. This mapping is valid as long as $f\gg1$ for typical momenta, with small enough diluteness parameter $\zeta$ according to (\ref{eq:overpopulation}). In classical-statistical simulations, one samples over initial conditions and evolves each realization according to the classical equation of motion (\ref{eq:gpe}). This equation is solved on a three-dimensional grid using a split-step method~\cite{Berges:2012us,Nowak:2011sk}. Then, observables are obtained by averaging over classical trajectories.

We concentrate on scaling properties of a time-dependent occupation number distribution $f(t,{\mbf p})$. More precisely, we consider the two-point correlation function 
\begin{equation}
F(t,t^\prime, \mbf x - \mbf x^\prime) = \frac{1}{2} \langle \psi(t,\mbf x)\psi^*(t^\prime,\mbf x^\prime) + \psi(t^\prime,\mbf x^\prime)\psi^*(t,\mbf x)\rangle
\label{eq:anticom}
\end{equation}
evaluated at equal times $t=t^\prime$ for spatially homogeneous ensembles. Brackets $\langle \ldots \rangle$ indicate sample averages. In spatial Fourier space we define~\cite{Berges:2007ym}: 
\begin{equation}
f(t,{\mbf p}) + (2 \pi)^3 \delta^{(3)}({\mbf p}) |\psi_0|^2(t) \equiv \int d^3x\, e^{-i {\mbf p} {\mbf x}}\, F(t,t, \mbf x) \, . \quad
\label{eq:n_def_stat_nonrel}
\end{equation} 
Because of spatial isotropy, the distribution function depends on the modulus of momentum, and we frequently write $f(t,|\mathbf p|)$. The term $\sim |\psi_0|^2(t)$ coming together with the Dirac \mbox{$\delta$-function} represents the condensate part of the correlator at zero momentum. The time-dependent condensate fraction at zero momentum is then given by 
\begin{equation}
\frac{N_0(t)}{N_{\rm total}}= \frac{|\psi_0|^2(t)}{\int d^3p/(2\pi)^3 f(t,{\mbf p}) + |\psi_0|^2(t)} \,.
\label{eq:condfrac} 
\end{equation}
The corresponding expressions for the finite volumes employed will be discussed in section~\ref{sec:condensationnonrel}.\footnote{In the quantum theory (\ref{eq:anticom}) denotes the anticommutator expectation value, which has a well-defined equal-time limit, and the definition of the distribution function is obtained by the substitution $f \rightarrow f + 1/2$ in (\ref{eq:n_def_stat_nonrel}). Since our typical occupation numbers are large, we drop here the `quantum-half'. Furthermore, 
for the class of initial conditions considered, no disconnected part of the correlation function arises. This does not exclude a nonzero condensate contribution that scales proportional to volume, which will be discussed in detail in section~\ref{sec:condensationnonrel}.\label{foot:qh}}

We consider initial conditions with high occupation numbers as motivated in the previous section. Specifically, we start with a distribution function of the form
\begin{equation}
 f(0,{\mbf p}) \sim \frac{1}{\zeta} \,\Theta(Q-|{\mbf p}|)\,,
 \label{eq:fluctuationIC-nonrel}
\end{equation}
which describes overoccupation up to the characteristic momentum $Q$. The initial condensate fraction is taken to be zero, i.e.~$|\psi_0|^2(t=0) = 0$, with an initial $f(0,{\mbf p}) = 50 / (2 m g Q) \,\Theta(Q-|{\mbf p}|)$. For the plots we typically average over 10 realizations on a lattice with $512^3$ points and a lattice spacing $a_s$ such that $Qa_s=1$, where we checked insensitivity of our infrared results to cutoff changes. We always plot dimensionless quantities obtained by the rescalings $f(t,{\mbf p})\rightarrow f(t,{\mbf p})\,2mgQ$, $t\rightarrow t\,Q^2/(2m)$ and ${\mbf p}\rightarrow {\mbf p}/Q$. This reflects the classical-statistical nature of the dynamics in the highly occupied regime, which has the important consequence that if we measure time in units of $2m/Q^2$ and momentum in units of $Q$ then the combination $f(t,{\mbf p})\,2mgQ$ does not depend on the values of $m$, $g$ and $Q$~\cite{Berges:2014xea}.

\begin{figure}[t!]
\centering
	\includegraphics[width=.5\textwidth]{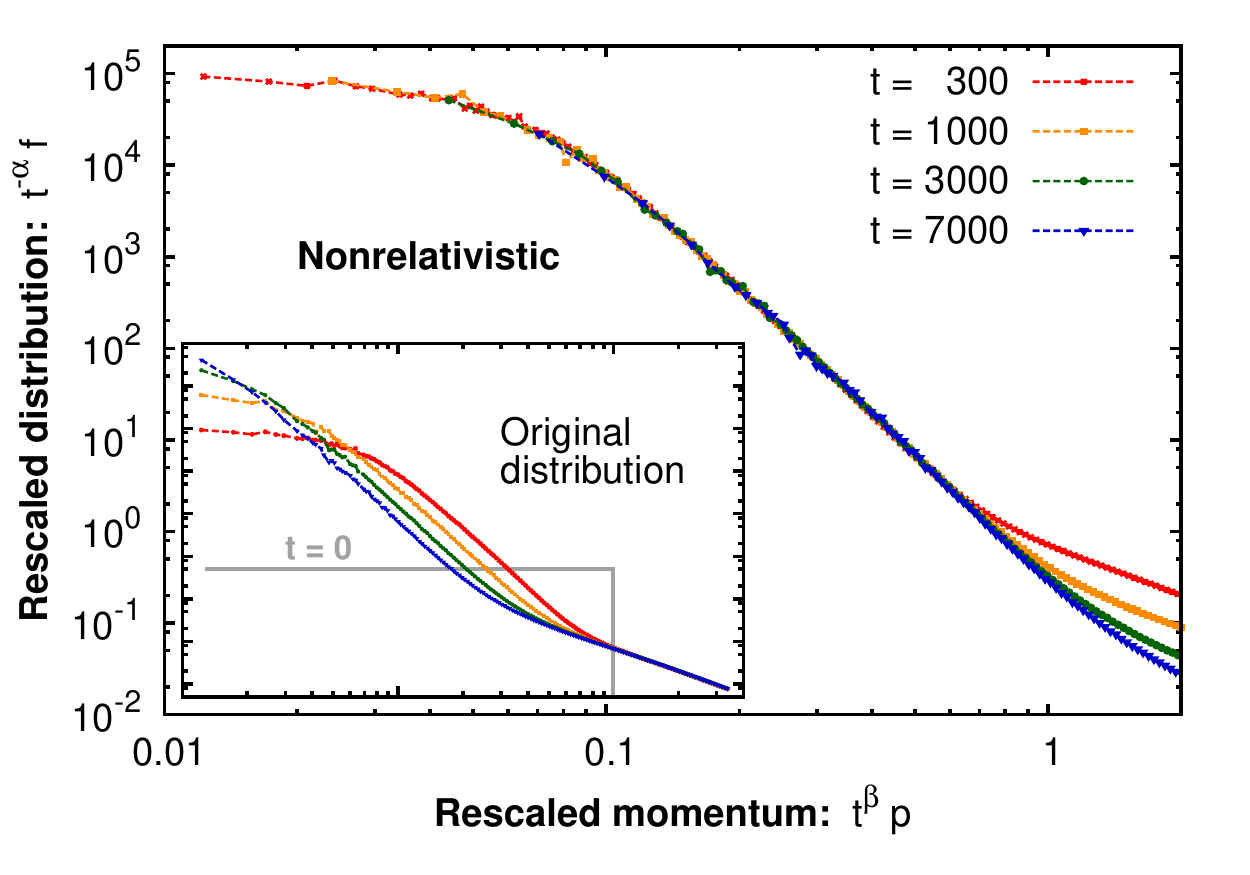}
	\caption{Rescaled distribution function of the nonrelativistic theory as a function of the rescaled momentum for different times. The inset shows the original distribution without rescaling.}
	\label{fig:ab_nonrel_1}
\end{figure}

The initial mode occupancies (\ref{eq:fluctuationIC-nonrel}) get quickly redistributed at the beginning of the nonequilibrium evolution and then a slower behavior sets in. The latter reflects the dynamics near the nonthermal fixed point, where universality can be observed. We concentrate on the low-momentum part of the distribution and analyze its infrared scaling properties. FIG.~\ref{fig:ab_nonrel_1} shows the rescaled distribution $(t/t_{\rm ref})^{-\alpha}f(t,{\mbf p})$ of the nonrelativistic theory as a function of $(t/t_{\rm ref})^\beta |{\mbf p}|$, where the reference time $t_{\rm ref} Q^2/(2m) = 300$ is the earliest time shown. The inset gives the curves at different times together with the initial distribution without rescaling for comparison. With the appropriate choice of the infrared scaling exponents $\alpha$ and $\beta$, all the curves at different times lie remarkably well on top of each other after rescaling. This is a striking manifestation of the self-similar dynamics (\ref{eq:selfsim}) near a nonthermal fixed point. The scaling exponents obtained are
\begin{align}
	\alpha=&\,1.66 \pm 0.12\,,\qquad \beta=\,0.55 \pm 0.03\,,
\label{eq:ab_results_nonrel}
\end{align}
where the error bars are due to statistical averaging and fitting errors, which is further described in appendix~\ref{app:error}. 

Comparing these values to (\ref{eq:exponentsIPC}), we find that the numerical results (\ref{eq:ab_results_nonrel}) agree rather well with the NLO approximation for a vanishing anomalous dimension, which are derived in section~\ref{sec:vrkt}. Furthermore, the simulation results confirm that $\alpha = 3\beta$ to very good accuracy as expected from number conservation in the infrared scaling regime (see section \ref{sec:intro}). The values for $\alpha$ and $\beta$ determine the rate and direction of the particle number transport, since according to (\ref{eq:selfsim}) a given characteristic momentum scale $K(t_0)=K_0$ evolves as $K(t)=K_0(t/t_0)^{-\beta}$ with amplitude $f(t,K(t))\sim t^\alpha$. Therefore, the positive values for the exponents in the infrared scaling regime imply that particle number is being transported towards low momenta, thus confirming an inverse particle cascade.

\subsection{Condensate formation}
\label{sec:condensationnonrel}

For the initial conditions (\ref{eq:fluctuationIC-nonrel}), there is no condensate present at $t=0$. However, the inverse particle cascade towards the infrared continuously populates the zero-mode, which leads to the formation of a condensate far from equilibrium~\cite{Svistunov:1,Kagan:1,Berloff:1,Berges:2012us,Nowak:2012gd,Berges:2014xea}. Near the nonthermal fixed point, the condensation dynamics is expected to exhibit scaling behavior and in the following we determine the relevant scaling exponent.

\begin{figure}[t!]
\centering
	\includegraphics[width=.5\textwidth]{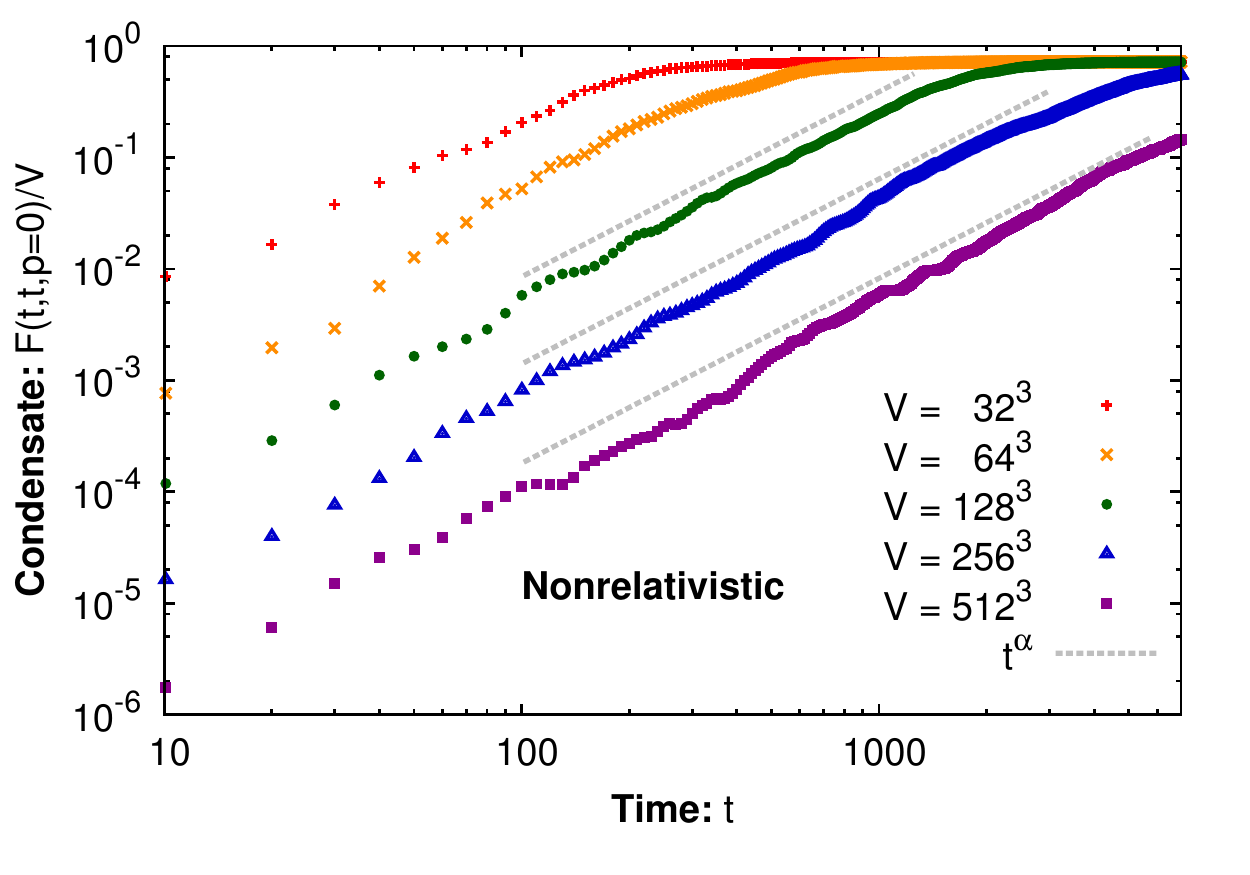}
	\caption{Evolution of the zero-momentum correlation divided by volume for the nonrelativistic Bose gas. A power-law $\sim t^{\alpha}$ has been fitted in the time interval with self-similar evolution for the three largest volumes (dashed lines). Subsequently, the results for different volumes converge, signaling the formation of a coherent zero mode spreading over the entire volume.}
	\label{fig:conden_build_nonrel}
\end{figure}

Our starting point is the Fourier transformed equal-time correlation function (\ref{eq:n_def_stat_nonrel}) with initially vanishing condensate fraction, $|\psi_0|^2(t=0) = 0$. To analyze our simulation results for $t>0$, we follow references~\cite{Berges:2012us,Berges:2014xea} and note that for a finite volume $V$ the zero-mode contribution in (\ref{eq:n_def_stat_nonrel}) scales with $(2\pi)^3\delta^{(3)}(\mbf 0) \rightarrow V$ proportional to volume. Therefore, we can identify the condensate fraction by its scaling behavior as the volume is changed. Stated differently, if we divide the correlation function (\ref{eq:n_def_stat_nonrel}) by the volume, only correlations which scale with the system size are not suppressed at large volumes and the condensate fraction is related to the volume-independent part. 

FIG.~\ref{fig:conden_build_nonrel} shows the evolution of the zero-momentum correlation $V^{-1} F(t,t,{\mbf p}=0)\equiv V^{-1} \int d^3x F(t,t,{\mathbf x})$ for different volumes. These are given in units of $Q$, ranging from $VQ^3 = 32^3$ to the largest volume $\sim 512^3$. Correspondingly, the plotted dimensionless results are rescaled as $V^{-1} F(t,t,{\mbf p}=0) \rightarrow V^{-1} F(t,t,{\mbf p}=0) \,2mgQ/Q^3$. One observes that at early times the evolution depends strongly on the volume, as expected in the absence of a coherent zero mode spreading over the entire volume. However, after a transient regime the zero-momentum correlation becomes volume independent. The curves corresponding to different volumes converge towards the same value, thus signaling the formation of a condensate. 

The double logarithmic plot of FIG.~\ref{fig:conden_build_nonrel} clearly indicates that the growth of the zero-momentum correlation proceeds as a power-law in time. The power-law growth is rather well described by the scaling exponent $\alpha$ found in (\ref{eq:ab_results_nonrel}) from the self-similar evolution of the distribution function. These numerical findings are also explained in more detail using the analytic scaling solution in section~\ref{sec:anomalousscaling}. 

\begin{figure}[t!]
\centering
	\includegraphics[width=.5\textwidth]{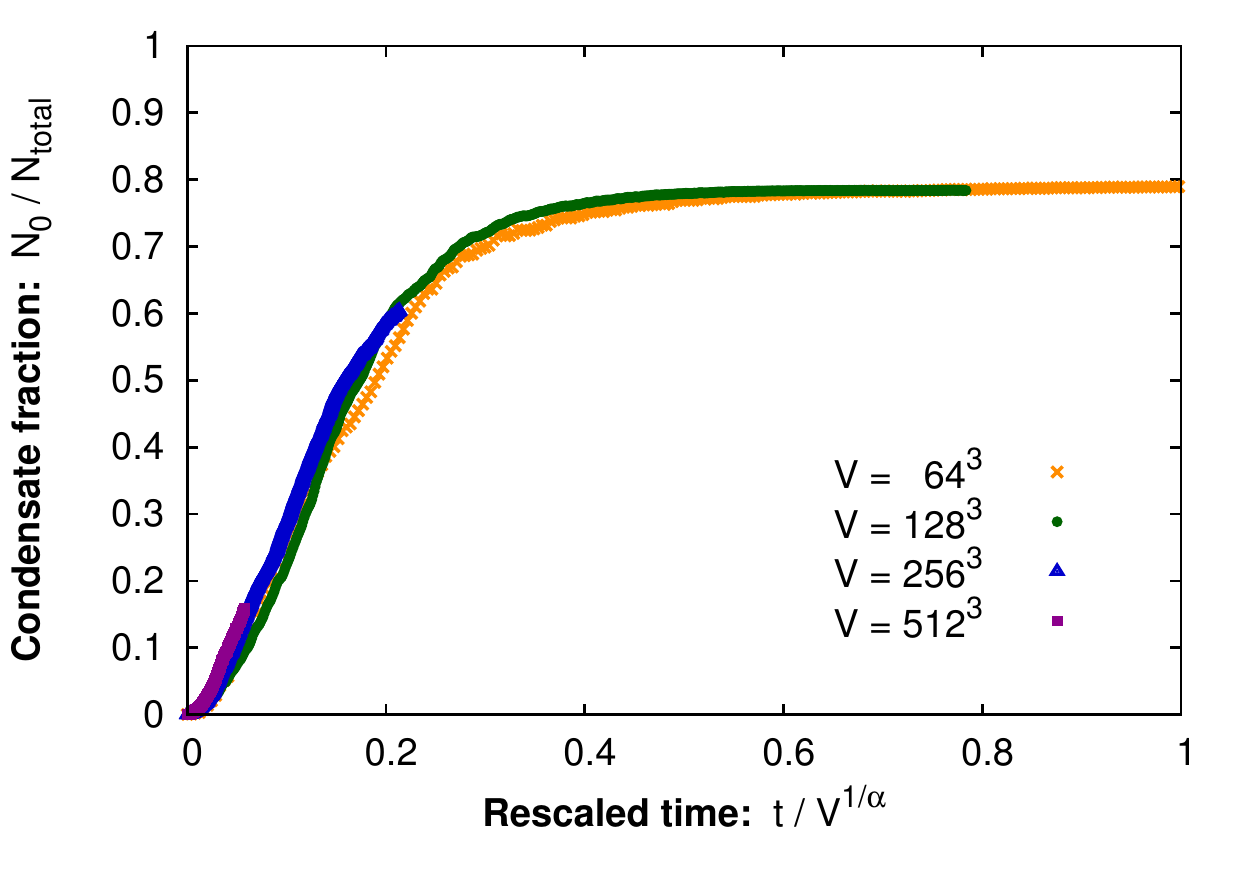}
	\caption{Evolution of the condensate fraction for the nonrelativistic Bose gas for different volumes $V$. The different curves become approximately volume independent after rescaling of time by $V^{-1/\alpha}$, in agreement with (\ref{eq:condens-time-estimate}).}
	\label{fig:conden_lin_nonrel}
\end{figure}

Of course, the time needed to fill the entire volume with a single condensate increases with volume, which is nicely observed from the data. Using that the parametrically slow power-law dynamics dominates the time until condensation is completed, we can use the scaling exponent $\alpha$ to estimate this condensation time. Taking the value of the zero-momentum correlator $V^{-1} F(t,t,{\mbf p}=0)$ at the initial time $t_0$ of the self-similar regime as $V^{-1}f(t_0,0)$ and its final value at the time $t_f$ as $|\psi_0|^2(t_f)$, we can estimate from $V^{-1} F(t,t,{\mbf p}=0) \sim t^\alpha$ the condensation time as 
\begin{equation}
t_f \,\simeq\,  t_0 \, \left(\frac{|\psi_0|^2(t_f)}{f(t_0,0)}\right)^{1/\alpha} \, V^{1/\alpha} \, .
\label{eq:condens-time-estimate}
\end{equation}
Correspondingly, we define the condensate fraction (\ref{eq:condfrac}) for the case of finite volumes as $N_0/N_{\rm total} \rightarrow V^{-1} F(t,t,{\mbf p}=0)/F(t,t,{\mathbf x}=0)$, using that $N_{\rm total} = F(t,t,{\mathbf x}=0)$. In Fig.~\ref{fig:conden_lin_nonrel} we show the evolution of the condensate fraction for different volumes as a function of the rescaled time $t/V^{1/\alpha}$. Indeed, as predicted by (\ref{eq:condens-time-estimate}), the different curves are approximately volume independent. One finds that the condensate fraction saturates at $N_0/N_{\rm total} \simeq 0.8$. 

\section{Relativistic scalar field theory\label{sec:rel}}

\subsection{Initial conditions}

Overoccupied relativistic quantum field theories play an important role in early-universe cosmology and high-energy collision experiments with heavy nuclei. For a large class of inflationary models of early-universe dynamics, the accelerated expansion of the universe during inflation leads to a large coherent field amplitude of the (scalar) inflaton. A subsequent decay via nonequilibrium instabilities can lead to highly occupied modes of the scalar field with a characteristic momentum $Q$~\cite{Greene:1997fu}. A similar overoccupation of modes can also occur in heavy-ion collisions at ultra-relativistic energies, where a longitudinally expanding plasma of highly occupied gluon fields is expected to form shortly after the collision~\cite{Gelis:2010nm}. For expanding systems there are even striking indications that the gluon plasma can belong to the same far-from-equilibrium universality class as a scalar field theory~\cite{Berges:2014bba}.

To be specific, we consider here $N$ real scalar fields $\varphi_a$ interacting via a weak quartic self-coupling $\lambda \ll 1$ in three spatial dimensions. The equations of motion for the $a=1,\ldots, N$ massless fields are  
\begin{align}
	\left( \partial_t^2 - \nabla^2 + \frac{\lambda}{6N}\varphi_a(t, \mbf x)\varphi_a(t, \mbf x) \right) \varphi_b(t, \mbf x)=&\,0 \, ,
\label{eq:kge}
\end{align}
where a sum over repeated indices is implied. The relevant two-point correlation function we denote as
\begin{equation}
F(t,t^\prime,\mbf x -\mbf x^\prime) = \frac{1}{2N} \langle \varphi_a(t,\mbf x) \varphi_a(t^\prime,\mbf x^\prime) + \varphi_a(t^\prime,\mbf x^\prime) \varphi_a(t,\mbf x) \rangle .
\label{eq:corrrel}
\end{equation}
At equal times $t= t^\prime$ this can be used to define a distribution function $f(t,{\mbf p})$ for the relativistic theory:\footnote{Similar to the nonrelativistic case, we drop here the `quantum-half' as explained in footnote \ref{foot:qh}.}
\begin{equation}
\frac{f(t,{\mbf p})}{\omega(t,{\mbf p})} + (2 \pi)^3 \delta^{(3)}({\mbf p}) \phi_0^2(t) \equiv \int d^3x\, e^{-i {\mbf p} {\mbf x}}\, F(t,t, \mbf x) \, .
\label{eq:reldeffcond}
\end{equation} 
This is in complete analogy to the definition (\ref{eq:n_def_stat_nonrel}) for the nonrelativistic system and we will refer to the term \mbox{$\sim \phi_0^2(t)$} as the condensate part. The only major difference is the appearance of the dispersion $\omega(t,{\mbf p})$ in the definition for the relativistic case, which is a consequence of the second-order differential equation in time for the fields (\ref{eq:kge}). 

Similar to the above discussion for the nonrelativistic system, we first characterize overoccupied initial conditions for the typical momentum Q. In a mean-field or large-$N$ approximation to leading order, one finds for the evolution equation of the correlation function (\ref{eq:corrrel}) in spatial Fourier space~(see e.g.~\cite{Berges:2004yj}):
\begin{equation}
\left( \partial_t^2 +{\mbf p}^2 + \frac{\lambda}{6} \int \frac{d^3q}{(2 \pi)^3} F(t,t,{\mbf q}) \right) F(t,t^\prime,{\mbf p}) = 0 \, . 
\label{eq:evolF}
\end{equation}
If there is no condensate initially, we can estimate parametrically the mean-field correction at sufficiently early times as
\begin{eqnarray}
&&\lambda \int d^3p\, F(t,t,{\mbf p}) \,\sim \, \lambda \int \frac{d^3p}{(2 \pi)^3} \frac{f(t,{\mbf p})}{\omega({\mbf p})} \nonumber\\
&& \sim \, \lambda \int^Q dp\, p^2 \frac{f(t,{\mbf p})}{|{\mbf p}|} \, \sim \,  \lambda f(t,Q)\, Q^2 \, , \quad
\end{eqnarray}
where we have taken a relativistic dispersion $\omega \sim |{\mbf p}|$ for massless particles.
One observes that this is of the same order as the typical kinetic energy term $\sim Q^2$ in (\ref{eq:evolF}) if the occupancy is as large as 
\begin{equation}
f(t,Q) \sim \frac{1}{\lambda} \gg 1\, .
\label{eq:reloo}
\end{equation}
Common scalar inflaton models for early universe dynamics have couplings of order $\lambda \sim 10^{-13}$, such that the typical occupancies are extremely large in that context. Comparing to (\ref{eq:overpopulation}), we note that the dimensionless self-coupling $\lambda$ plays the role of the diluteness parameter $\zeta$ in the nonrelativistic theory.

\subsection{Self-similar dynamics from classical-statistical simulations}
\label{sec:simreldyn}

Since the self-similar dynamics can only be observed beyond the mean-field approximation, we perform first classical-statistical lattice simulations similar to what is done for the nonrelativistic theory in section~\ref{sec:simcl}. For this we solve~(\ref{eq:kge}) as classical equations of motion on a three dimensional lattice using a leapfrog algorithm~\cite{Berges:2008wm,Berges:2013lsa} and sample over initial conditions. We extract the occupation number distribution according to (\ref{eq:reldeffcond}) by writing for \mbox{$|{\mbf p}| > 0$}:
\begin{equation}
\frac{f(t,{\mbf p})}{\omega(t,{\mbf p})} = F(t,t,{\mbf p}) = \frac{\sqrt{F(t,t',{\mbf p}) \partial_{t}\partial_{t'}F(t,t',{\mbf p})}}{\sqrt{\partial_{t}\partial_{t'}F(t,t',{\mbf p})/F(t,t',{\mbf p})}}\Big|_{t=t'}\,.
\end{equation}
The second equality allows us to identify the dimensionless distribution~(see e.g.~\cite{Berges:2004yj})
\begin{align}
	f(t,{\mbf p})=&\, \sqrt{F(t,t',{\mbf p})\partial_{t}\partial_{t'}F(t,t',{\mbf p})}\,\Big|_{t=t'}\,.
\label{eq:n_def_stat_rel}
\end{align}
The corresponding dispersion relation is then given by
\begin{align}
	\omega(t,{\mbf p})=&\, \sqrt{\frac{\partial_{t}\partial_{t'}F(t,t',{\mbf p})}{F(t,t',{\mbf p})}}\,\Bigg|_{t=t'} \, .
\label{eq:disp_rel_stat}
\end{align}
We emphasize that the notion of particle number is not a uniquely defined concept in a relativistic field theory, where total number changing processes are possible. However, one may always think in terms of the well-defined correlation functions appearing in (\ref{eq:n_def_stat_rel}). Moreover, this definition turns out to provide an extremely useful quasi-particle interpretation even in the strongly correlated infrared regime~\cite{Berges:2008wm}, which we will exploit further in section (\ref{sec:vrkt}). 

Similar to (\ref{eq:fluctuationIC-nonrel}) for the nonrelativistic theory, we choose overoccupied initial conditions 
\begin{equation}
 f(0,{\mbf p}) \sim \frac{1}{\lambda} \,\Theta(Q-|\mbf p|)
 \label{eq:fluctuationIC-rel}
\end{equation}
with $\phi_0^2 (t=0) = 0$. All quantities shown will be made dimensionless by appropriate powers of the scale $Q_\epsilon = \sqrt[4]{\lambda \epsilon/N}$, which is obtained from the conserved energy density average
\begin{align}
 \epsilon = \left\langle \frac{1}{2} (\partial_t \varphi_a)(\partial_t \varphi_a) + \frac{1}{2}(\partial_i \varphi_a)(\partial_i \varphi_a) + \frac{\lambda}{4! N} (\varphi_a\varphi_a)^2 \right\rangle\;,
\end{align}
where summation over spatial directions $i$ and scalar components $a$ is implied. For the initial condition (\ref{eq:fluctuationIC-rel}), one has $\epsilon \sim N Q^4/\lambda$ such that $Q_\epsilon$ becomes independent of the coupling and the number of field components. 

For the figures we have chosen the amplitude of (\ref{eq:fluctuationIC-rel}) as $f(0,{\mbf p}) = 125/\lambda \,\Theta(Q-|{\mbf p}|)$ with $Q = 0.8\,Q_\epsilon$, and we always plot rescaled functions $f(t,{\mbf p})\rightarrow \lambda\,f(t,{\mbf p})$ and $F(t,t,{\mbf p})\rightarrow \lambda\,F(t,t,{\mbf p})$ such that these combinations also become independent of the coupling. For the $N = 2$ component theory, computations were made on a $768^3$ lattice with lattice spacing $Q_\epsilon a_s = 0.9$ and we averaged over five realizations. For $N=4$ we employed a $512^3$ lattice with spacing $Q_\epsilon a_s = 1.8$ and averaged over $18  - 50$ realizations. We checked that for $N=2$ all shown results are insensitive to the lattice spacing and the volume. For the relevant infrared quantities this is to good accuracy also the case for the coarser lattices employed for $N=4$.  

\begin{figure}[t!]
\centering
	\includegraphics[width=.5\textwidth]{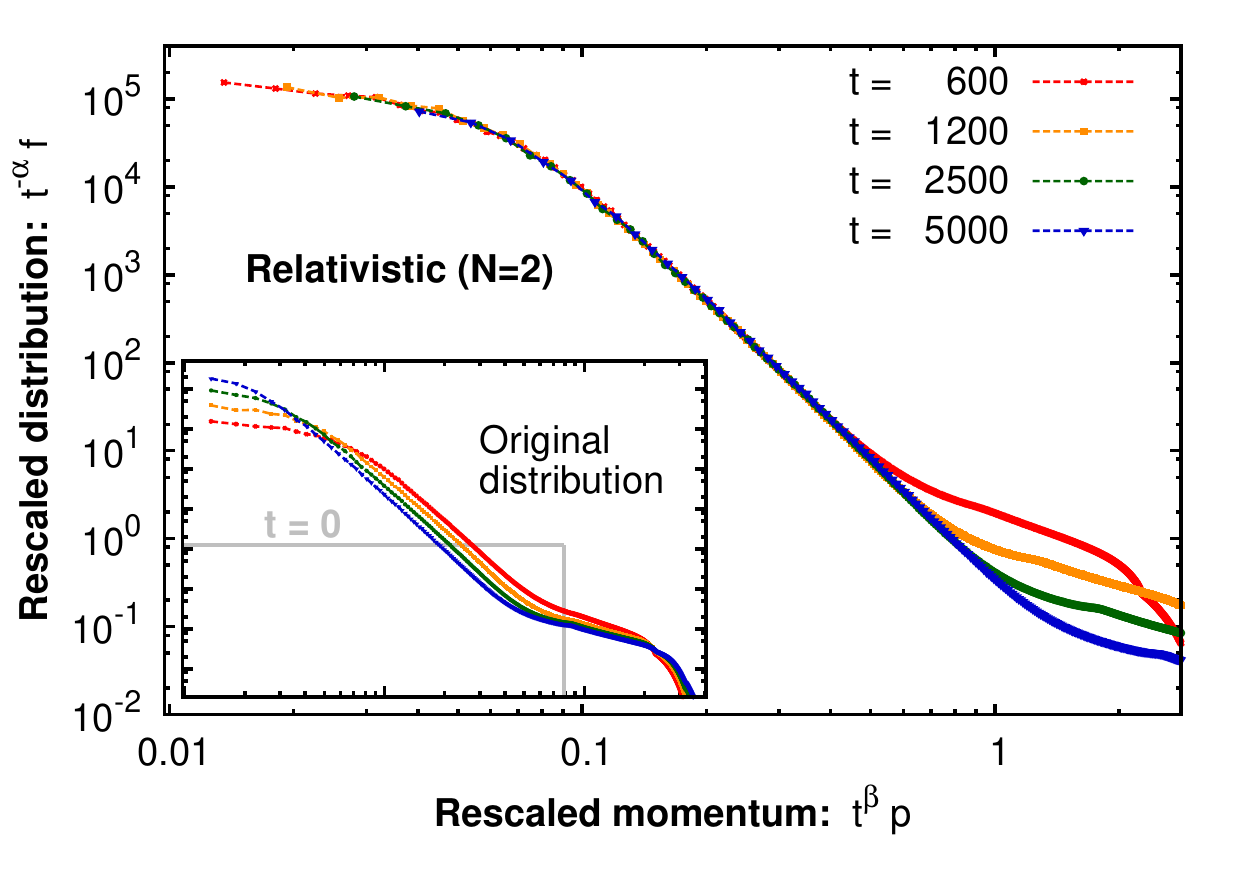}
	\caption{Rescaled distribution function of the relativistic two-component theory as a function of the rescaled momentum for different times. The inset shows the original distribution without rescaling.}
	\label{fig:rel_o2_ab_box}
\end{figure}

In FIG.~\ref{fig:rel_o2_ab_box} we show the evolution of the occupation number distribution for the relativistic $N=2$ component theory. As in the nonrelativistic case, we plot $t^{-\alpha} f$ against $t^\beta |{\mbf p}|$ to study self-similarity and give the original distribution without rescalings in the inset. With appropriately chosen exponents, in the infrared all curves lie on top of each other after rescaling to remarkable accuracy. The measured exponents are
\begin{align}
	\alpha=&\,1.51 \pm 0.13\,,\qquad \beta=\,0.51 \pm 0.04\, ,
\label{eq:ab_result_rel}
\end{align}
and we refer to appendix~{\ref{app:error}} for details on how we estimate the error bars. 

\begin{figure}[t!]
\centering
	\includegraphics[width=.5\textwidth]{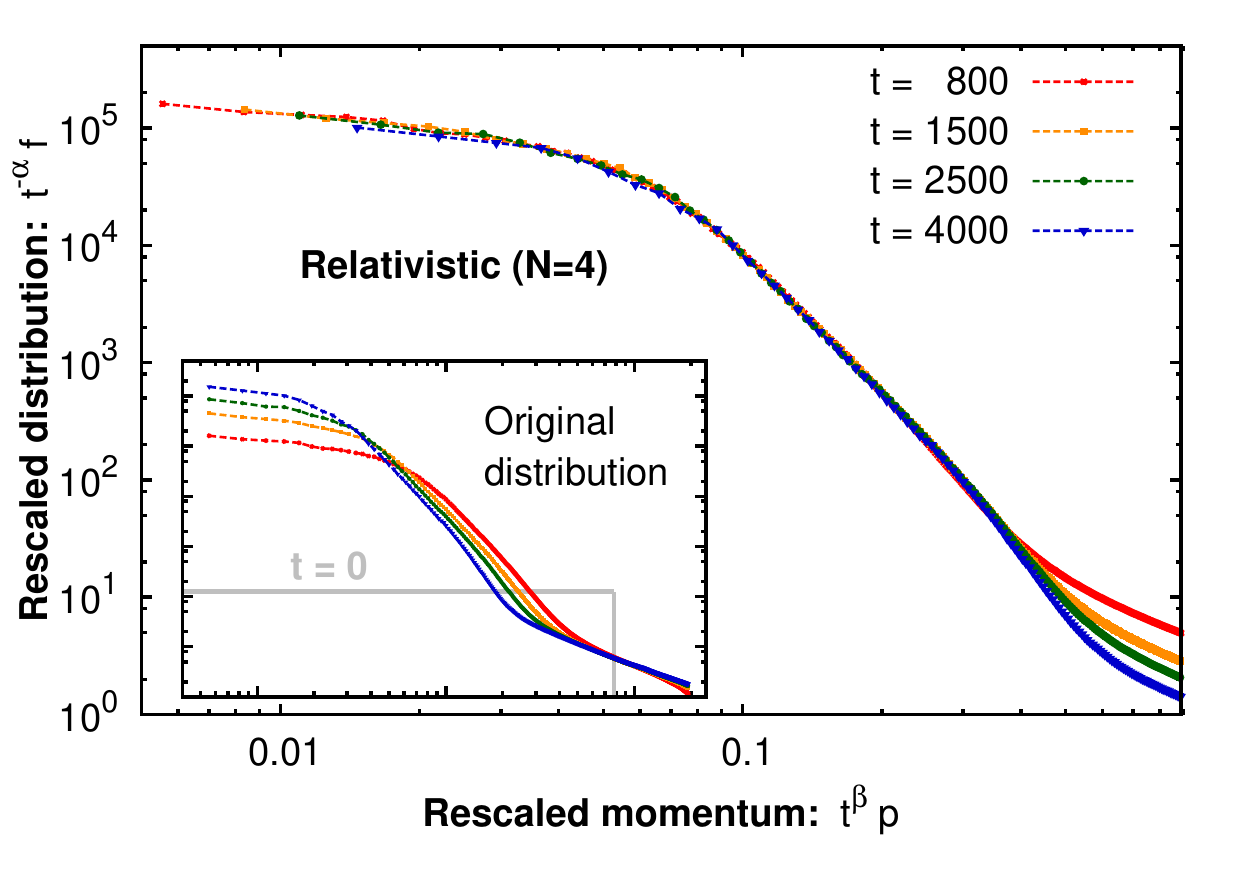}
	\caption{Rescaled distribution function of the relativistic four-component theory as a function of the rescaled momentum for different times. The inset shows the original distribution without rescaling.}
	\label{fig:rel_o4_ab_box}
\end{figure}

In order to check for a possible dependence of the infrared scaling properties on the number of field components, we also perform lattice simulations for $N=4$. FIG.~\ref{fig:rel_o4_ab_box} shows $t^{-\alpha} f$ as a function $t^\beta |{\mbf p}|$. The curves corresponding to different times lie again on top of each other after the rescalings and we extract the exponents 
\begin{align}
 \alpha=&\,1.65 \pm 0.09\,,\qquad \beta=\,0.59 \pm 0.03\,.
\label{eq:expN4}
\end{align}
These and the universal scaling form of the distribution function compare rather well to those for the relativistic $N=2$ component system as well as the nonrelativistic theory. Within the statistical errors we find no indication for a dependence of the corresponding universality class on the number of field components.  However, small discrepancies in $\alpha$, $\beta$ and $f_S$ are still possible. These could occur in the presence of nonvanishing anomalous dimension $\eta$, which is discussed further in section \ref{sec:anomalousscaling}. 

\begin{figure}[t!]
\centering
	\includegraphics[width=.5\textwidth]{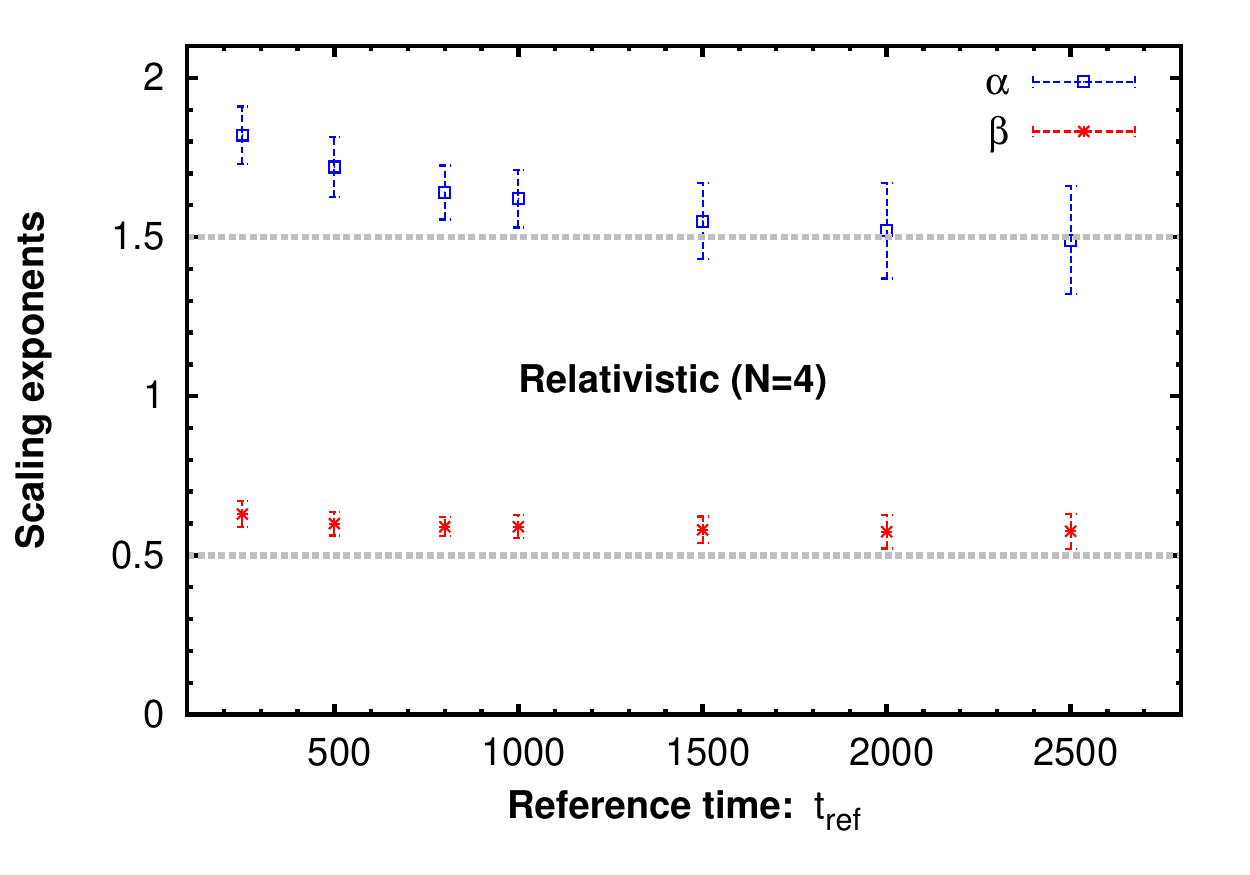}
	\caption{The exponents $\alpha$ and $\beta$ with statistical error bars for the relativistic $N=4$ component field theory extracted at different reference times $t_{\text{ref}}$.}
	\label{fig:LikelihoodATime82400V3}
\end{figure}

In order to estimate systematic errors, we investigate how the values for the exponents $\alpha$ and $\beta$ depend on the reference time $t_{\text{ref}}$ at which we start our self-similarity analysis. To this end, we perform our analysis for different values of $t_{\text{ref}}$ and use the method of appendix~\ref{app:error} with the distribution function evaluated at several times up to $t/t_{\text{ref}} \lesssim 4 - 5$. In FIG.~\ref{fig:LikelihoodATime82400V3}, we show the extracted values for $\alpha$ and $\beta$ as a function of the reference time for $N=4$. One finds that the mean value of $\alpha$ decreases monotonically to about $\alpha \approx 1.5$ while $\beta$ gets close to a half for the transient times at which self-similarity can be accurately observed. For the relativistic two-component system, the exponents $\alpha$ and $\beta$ are found to start from comparably larger values at early reference times to the ones given in~(\ref{eq:ab_result_rel}). We note that the nonrelativistic system of section~\ref{sec:simcl} also shows decreasing exponents $\alpha$ and $\beta$ but our runs last not long enough such that their mean values (\ref{eq:ab_results_nonrel}) come as close to the expected values $\alpha=1.5$ and $\beta=0.5$ as for the relativistic two-component system. 

\subsection{Generation of a mass gap}
\label{sec:massgap}

The values (\ref{eq:ab_result_rel}) or (\ref{eq:expN4}) for the exponents of the relativistic theories agree within errors with the values (\ref{eq:ab_results_nonrel}) obtained for the nonrelativistic Bose gas. Moreover, they are rather close to the analytic results (\ref{eq:exponentsIPC}) from the large-$N$ expansion to NLO for the nonrelativistic theory, while they deviate clearly from the corresponding prediction assuming a relativistic dispersion as will be shown in section~\ref{sec:vrkt}. 

Also the universal form of the nonthermal fixed point distribution $f_S$ accurately agrees for relativistic and nonrelativistic theories in the infrared scaling regime. This is shown in FIG.~\ref{fig:compare_rel_nonrel} by comparing the two-component relativistic and the nonrelativistic theory, where $f_S/A$ is given as a function of $t^\beta |{\mbf p}|/B$ normalized to the nonuniversal amplitudes $A$ and $B$ as described in section~\ref{sec:outline}.    

\begin{figure}[t!]
\centering
	\includegraphics[width=.5\textwidth]{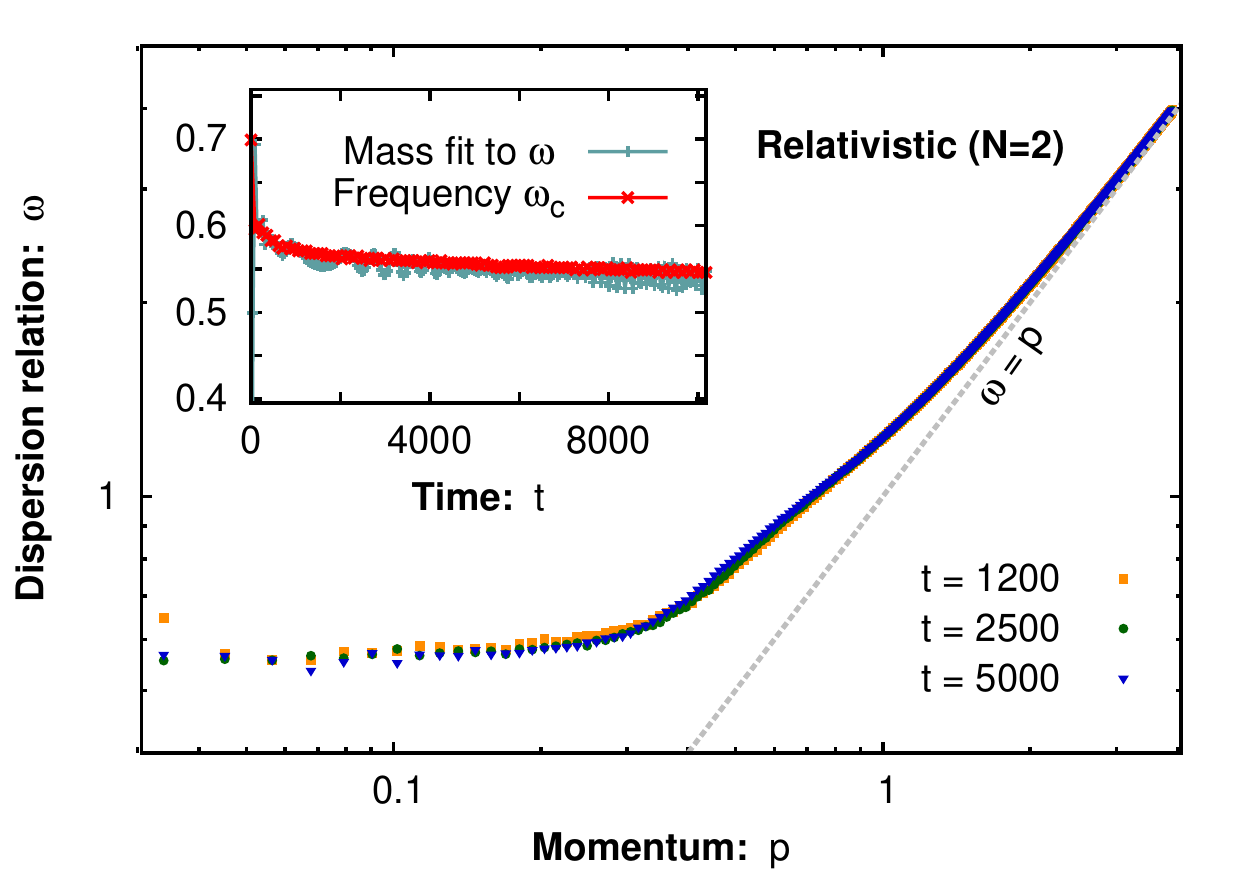}
	\caption{Dispersion relation for the relativistic $N=2$ component theory at different times for the same parameters as for FIG.~\ref{fig:rel_o2_ab_box}. In the inset, the time evolution of the mass gap at zero momentum is shown, which is obtained from a $\sqrt{m^2 + {\mbf p}^2}$ fit to $\omega_{{\mbf p}}$ at low momenta  (+ symbol) and by measuring the oscillation frequency $\omega_c$ of the unequal-time correlation function (x symbol).}
	\label{fig:disprel_o2}
\end{figure}

To understand the appearance of nonrelativistic dynamics, we analyze the dispersion relation $\omega(t,{\mbf p})$ according to (\ref{eq:disp_rel_stat}),
which is shown in FIG.~\ref{fig:disprel_o2} at three different times. Although the underlying theory is massless, it can be clearly observed that for low momenta the system generates a mass gap, whereas at large momenta we recover a linear dispersion. 

The appearance of an effective mass-like contribution can already be understood qualitatively from the approximate evolution equation (\ref{eq:evolF}) for the correlator modes $F(t,t^\prime,{\mbf p})$. In the mean-field approximation, the term $\sim (\lambda/6) \int d^3q/(2 \pi)^3 F(t,t,{\mbf q})$ generates a mass-like correction for the overoccupied initial condition (\ref{eq:fluctuationIC-rel}). 

To extract the mass gap beyond the mean-field approximation using the lattice simulations, we fit a time-dependent effective mass $m(t)$ from a $\sqrt{m^2(t)+{\mbf p}^2}$ fit to the $\omega(t,{\mbf p})$ data. The time evolution of $m(t)$ is shown in the inset of FIG.~\ref{fig:disprel_o2}. We find that after a quick initial evolution this dispersion relation enters a quasi-stationary regime, which is typical for a prethermalized quantity whose transient evolution is governed by an approximately conserved (particle) number~\cite{Berges:2004ce}.

In addition, we also analyze the oscillation frequency $\omega_c$ of the unequal-time correlation function $F(t,0,{\mbf p})$ as a function of time $t$ for $|{\mbf p}|=0$. Since the zero-momentum mode frequency corresponds to the renormalized mass of the theory, this provides an independent estimate of the mass gap that does not rely on the definition (\ref{eq:disp_rel_stat}) of a dispersion. Indeed we find $\omega_c \simeq m$ to very good accuracy as shown in the inset of FIG.~\ref{fig:disprel_o2}.    

In the presence of a mass gap $m$, low momentum modes with $|{\mbf p}| \lesssim m$ are expected to behave nonrelativistically. From FIG.~\ref{fig:disprel_o2} we can estimate the mass to be $m \simeq 0.55 Q_\epsilon \simeq 0.69 Q$ for the whole duration of the self-similar evolution. We find that this mass scale separates rather well the inertial ranges for the inverse particle cascade towards low momenta from the direct energy cascade at higher momenta. This can be observed, for instance, from the inset of FIG.~\ref{fig:rel_o2_ab_box}, where the initial scale $Q$ marked by the distribution at $t=0$ can be used as a reference. 

The emergence of a mass gap in the relativistic theory explains why the dynamics in the infrared regime is governed by nonrelativistic physics. Of course, in general the presence of a mass gap does not necessarily imply universal behavior for sufficiently low momentum modes. However, it may be seen as a prerequisite for relativistic theories to belong to the same far-from-equilibrium universality class as the Gross-Pitaevskii field theory.

\subsection{Condensate formation}
\label{sec:relcond}

Since we observe for the relativistic scalar field theory the same inverse particle cascade with universal exponents and scaling function as for the nonrelativistic system, one expects condensation to proceed with the same exponent $\alpha$ as found in section \ref{sec:condensationnonrel}. However, this is not entirely trivial since scatterings off the condensate play an important role in the inertial range for the direct energy cascade towards higher momenta, where both theories do not belong to the same universality class~\cite{Micha:2002ey,Micha:2004bv}.
  
\begin{figure}[t!]
\centering
	\includegraphics[width=.5\textwidth]{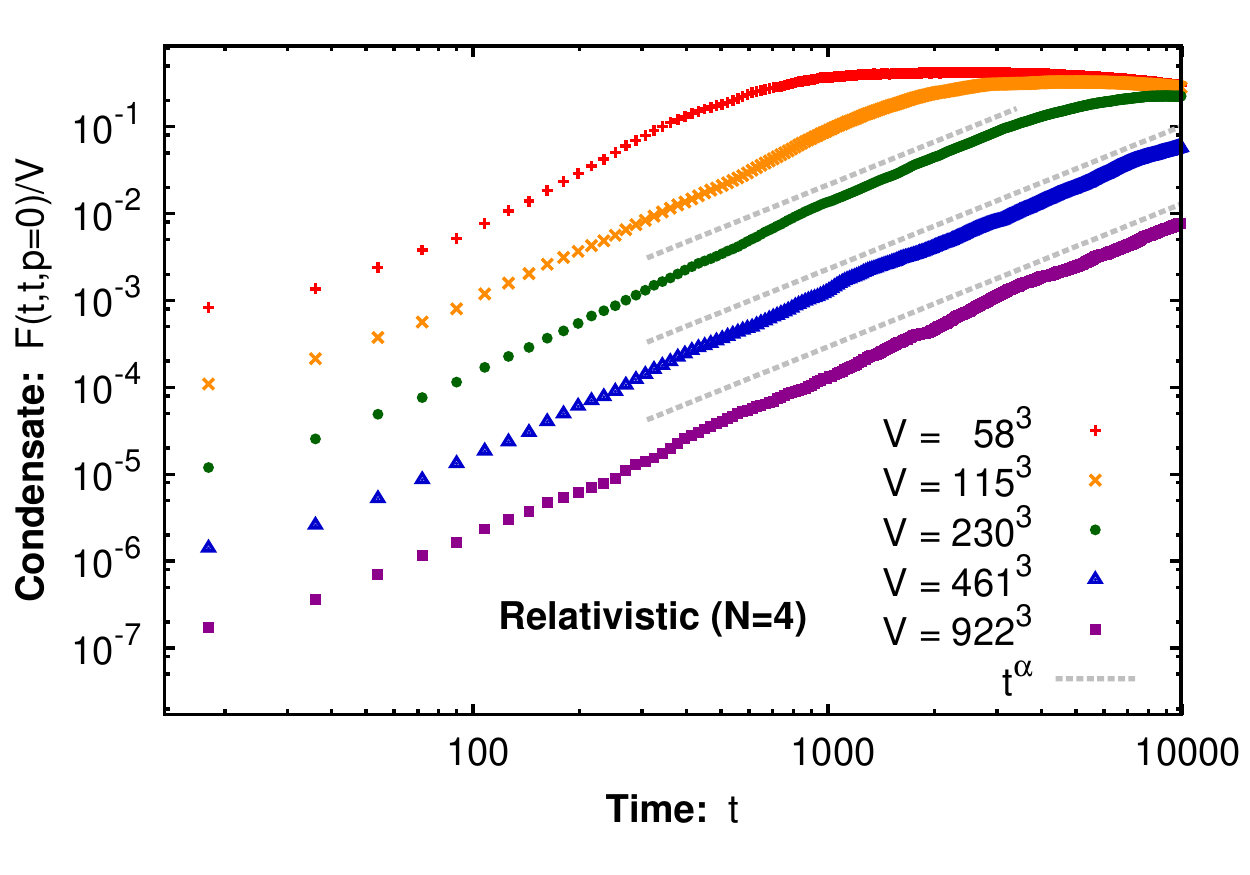}
	\caption{Evolution of the zero-momentum correlation divided by volume for the relativistic $N=4$ component theory, similar to FIG.~\ref{fig:conden_build_nonrel} for the nonrelativistic case.}
	\label{fig:conden_build_rel}
\end{figure}

In order to clarify this, we analyze the growth of the condensate during the self-similar regime for the $N=4$ component theory along the same lines as in section~\ref{sec:condensationnonrel}. The results are given in FIG.~\ref{fig:conden_build_rel} for different volumes, ranging from $VQ_\epsilon^3 = 58^3$ to the largest volume $\sim 922^3$.
Shown is the time evolution of the zero-mode of the correlation function (\ref{eq:reldeffcond}) divided by volume. We compare the curves to the expected power-law behavior $\sim t^\alpha$ with the exponent $\alpha$ obtained before from the self-similarity analysis. As in the nonrelativistic case, the dynamics is well described by such a power-law until the results become volume independent, thus signaling the formation of a coherent condensate $\phi_0^2$ for the entire volume. 

We emphasize that the observed power-law is restricted to the formation of the condensate during the transient self-similar regime, where we find the universal exponent $\alpha$ to govern the dynamics. In particular, we do not discuss here the subsequent late-time approach to thermal equilibrium, where total particle number changing processes in the relativistic theory can make important differences as compared to the number conserving nonrelativistic system~\cite{Berges:2012us}.

\section{Nonthermal fixed points from vertex-resummed kinetic theory}
\label{sec:vrkt}

\subsection{Self-similarity}

While many aspects of transport of conserved charges associated to wave turbulence in kinetic theory have long reached textbook level, there is still rather little known about the analytic description of turbulent behavior in nonperturbative regimes of quantum field theories. 

Perturbative kinetic theory has been successfully applied to the phenomenon of weak wave turbulence~\cite{Zakharov:1992,Nazarenko:2011,Micha:2004bv}. It has also been employed in the literature to describe infrared phenomena such as Bose condensation~\cite{Semikoz:1994zp,Semikoz:1995rd}. However, in this case the approach neglects important vertex corrections since the large occupancies at low momenta lead to strongly nonlinear dynamics~\cite{Berges:2008wm,Nowak:2011sk}. Here we apply a vertex-resummed kinetic theory, which is based on an expansion in the number of field components $N$ to next-to-leading order~\cite{Aarts:2002dj,Berges:2001fi,Scheppach:2009wu}. It extends well-known kinetic descriptions~\cite{Zakharov:1992,Nazarenko:2011,Micha:2004bv} to the nonperturbative regime of overoccupied modes. This allows us to gain analytic understanding of the formation of a dual cascade, in which particles are also transferred towards low momenta leading to Bose condensation. 

Similar nonperturbative descriptions have been employed before for the scaling behavior of stationary transport, which is time-translation invariant~\cite{Berges:2008wm,Berges:2008sr,Scheppach:2009wu}. Since our isolated systems out of equilibrium cannot realize stationary transport solutions in the absence of external driving forces, we have to consider here the more general notion of a self-similar evolution. Though this is not time-translation invariant, the dynamics in this case is described by time-independent scaling functions and scaling exponents. 

A self-similar evolution of the distribution function $f(t,\mbf p)$ for a spatially homogeneous and isotropic system is characterized as 
\begin{equation}
f(t,\mbf p)=s^{\alpha/\beta} f(s^{-1/\beta}t,s \mbf p) \, 
\label{eq:gen_selfsim_ansatz}
\end{equation}
with the real scaling parameter $s$ and exponents $\alpha$ and $\beta$. Again, all quantities are considered to be dimensionless by use of some suitable momentum scale. Choosing $s^{-1/\beta}t = 1$ we recover (\ref{eq:selfsim}), where the time-independent scaling function $f_S(t^\beta \mbf p)\equiv f(1,t^\beta \mbf p)$ denotes the {\it fixed point distribution}. This scaling form represents an enormous reduction of the possible dependence of the dynamics on variations in time and momenta, since $t^{-\alpha}f(t,\mbf p)$ depends on the product $t^\beta |\mbf p|$ instead of separately depending on time and momenta. Therefore, an essential part of the time evolution is encoded in the momentum dependence of the fixed point distribution $f_S(\mbf p)$.

For the time evolution of the distribution function $f(t,\mbf p)$ we write
\begin{equation}
\frac{\partial f(t,\mbf p)}{\partial t} = C\left[f\right](t,\mbf p)
\label{eq:genkineq}
\end{equation}
with a generic `collision integral' $C\left[f\right](t,\mbf p)$, which depends on the theory and the approximation employed. For the self-similar distribution (\ref{eq:gen_selfsim_ansatz}), the scaling behavior of the collision integral is then given by 
\begin{equation}
C[f](t,\mbf p) = s^{-\mu}\, C[f](s^{-1/\beta}t,s \mbf p) = t^{-\beta \mu}\, C[f_S](1,t^\beta \mbf p) \, ,
\label{eq:selfsimC}
\end{equation}
where $\mu$ is a function of scaling exponents such as $\alpha$ and $\beta$. Substituting this scaling form into the kinetic equation leads to the time-independent {\it fixed point equation} for $f_S(\mbf p)$:
\begin{equation}
\left[ \alpha+\beta\,\mbf p \cdot\mathbf{\nabla}_{\mbf p} \right] f_S(\mbf p) = C[f_S](1,\mbf p) \, , \label{eq:fpequation}
\end{equation}
and the {\it scaling relation}:
\begin{equation}
\alpha - 1 = - \beta \mu \, .
\label{eq:scalingrelab}
\end{equation}
This follows from comparing the LHS of the kinetic equation,
\begin{equation}
\frac{\partial}{\partial t}\left[ t^\alpha\, f_S(t^\beta \mbf p) \right]
	= t^{\alpha-1}\left[ \alpha+\beta\,{\mbf q}\cdot\nabla_{\mbf q} \right] f_S(\mbf q) |_{\mbf q = 
	t^\beta \mbf p} \, ,
\label{eq:lhs_pert}
\end{equation}
to its RHS given by (\ref{eq:selfsimC}).

Further relations can be obtained by either imposing energy conservation or particle number conservation if applicable.
For constant  
\begin{equation}
 n = \int \frac{d^dp}{(2\pi)^d}\, f(t,\mbf p) = t^{\alpha - \beta d} \int \frac{d^dq}{(2\pi)^d}\, f_S(\mbf q)
\end{equation}
one obtains the relation for 
\begin{equation} 
\!\!\!\!\!\!\mbox{\it particle conservation:} \quad  \alpha = \beta d \, .
\label{eq:pcrel}
\end{equation}
Similarly, with
\begin{equation}
\epsilon = \int \frac{d^dp}{(2\pi)^d}\, \omega(\mbf p) f(t,\mbf p) = t^{\alpha - \beta (d+z)} \int \frac{d^dq}{(2\pi)^d}\, \omega(\mbf q) f_S(\mbf q)
\end{equation}
one obtains from
\begin{equation} 
\!\!\!\!\!\!\mbox{\it energy conservation:} \quad  \alpha = \beta (d+z) \, ,
\label{eq:ecrel}
\end{equation}
where the dispersion is taken to scale with the dynamic exponent $z$ as
\begin{equation}
\omega(\mbf p) = s^{-z} \omega(s \mbf p) \, 
\label{eq:zdisp}
\end{equation}
with $z=1$ for the gapless relativistic and $z=2$ for the nonrelativistic theory. 
One observes that there is no single scaling solution conserving both energy and particle number. As outlined already above, in this case a dual cascade is expected to emerge such that in a given inertial range of momenta only one conservation law governs the scaling behavior.

\subsection{Perturbative scaling behavior}
\label{sec:perturbativescaling}

We first review some perturbative results for later comparison.
In perturbative kinetic theory, when two particles scatter into two particles, the collision integral is of the form
\begin{eqnarray} 
&& C^{2\leftrightarrow 2}[f](t,{\mbf p}) = \int {\mathrm{d}}\Omega^{2\leftrightarrow 2}({\mbf p},{\mbf l},{\mbf q},{\mbf r})  \nonumber\\
&& \quad \times\left[(f_{\mbf p}+1) (f_{\mbf l}+1) f_{\mbf q} f_{\mbf r} - f_{\mbf p} f_{\mbf l} (f_{\mbf q}+1) (f_{\mbf r}+1)\right] , \quad
\label{eq:2to2}
\end{eqnarray}
where we write $f(t,\mbf p) \equiv f_{\mbf p}$ suppressing the global time dependence to ease the notation. The details of the model enter $\int {\mathrm{d}}\Omega^{2\leftrightarrow 2}({\mbf p},{\mbf l},{\mbf q},{\mbf r})$, which for the example of the relativistic $N$-component scalar field theory with quartic $\lambda/(4!N)$-interaction of section~\ref{sec:rel} reads: 
\begin{eqnarray} 
\int d\Omega^{2\leftrightarrow 2}({\mbf p},{\mbf l},{\mbf q},{\mbf r}) =  \lambda^2\, \frac{N+2}{6 N^2} \int\frac{d^d l}{(2\pi)^d}\frac{d^d q}{(2\pi)^d}\frac{d^d r}{(2\pi)^d} && \nonumber\\
\times(2\pi)^{d+1}\, \delta^{(d)}({\mbf p}+{\mbf l}-{\mbf q}-{\mbf r})\, 
\frac{\delta(\omega_{\mbf p} + \omega_{\mbf l} - \omega_{\mbf q} - \omega_{\mbf r})}{2\omega_{\mbf p} 2\omega_{\mbf l} 2\omega_{\mbf q} 2\omega_{\mbf r}} &&  \nonumber\\
\label{eq:measurepert}
\end{eqnarray}
with $\omega_{\mbf p} = \sqrt{{\mbf p}^2 + m^2}$. 

The expression represents a standard Boltzmann equation for a gas of relativistic particles, which 
is not expected to be valid if the occupation numbers per mode become too large. Parametrically, for a weak coupling $\lambda$ a necessary condition for its validity is $f_{\mbf p} \ll 1/\lambda$ since otherwise strongly nonlinear effects become significant as will be explained in detail in section~\ref{sec:npscaling}. On the other hand, scaling is expected for not too small occupation numbers per mode, which we discuss now. For the corresponding regime $1 \ll f_{\mbf p} \ll 1/\lambda$ one may use the above Boltzmann equation, which approximately becomes 
\begin{eqnarray} 
\frac{\partial }{\partial t} f_{\mbf p} &\simeq& \int {\mathrm{d}}\Omega^{2\leftrightarrow 2}({\mbf p},{\mbf l},{\mbf q},{\mbf r})  \nonumber\\
&& \times \left[ 
(f_{\mbf p} + f_{\mbf l}) f_{\mbf q} f_{\mbf r} - f_{\mbf p} f_{\mbf l} (f_{\mbf q} + f_{\mbf r})\right]  \, .
\label{eq:2to2cl}
\end{eqnarray}
Apart from the energy density $\epsilon$ also the total particle number density $n$ is conserved by the collision term.

For the relativistic theory, the scaling assumption (\ref{eq:gen_selfsim_ansatz}) should be valid for sufficiently high momenta $|{\mbf p}| \gg m$ such that the dispersion is approximately linear with $\omega_{\mbf p} \sim |{\mbf p}|$. In this case one obtains for the scaling of the collision integral (\ref{eq:measurepert}) and (\ref{eq:2to2cl}) of the theory with quartic self-interaction:
\begin{equation} 
C^{2\leftrightarrow 2}[f](t,{\mbf p}) \, = \, s^{-\mu_4}\, C^{2\leftrightarrow 2}[f](s^{-1/\beta}t,s \mbf p) \, , 
\end{equation}
where the scaling is described by 
\begin{equation} 
\mu_4 \, = \, (3d-4)-(d+1) - 3\alpha/\beta\, = \, 2d - 5 - 3\alpha/\beta\, .
\label{eq:Del4}
\end{equation}
The first term in brackets comes from the scaling of the integral measure, the second from energy-momentum conservation for two-to-two scattering and the third from the three factors of the distribution function appearing in (\ref{eq:2to2cl}). 

Apart from the 4-vertex interaction considered, it will be relevant to investigate also scattering in the presence of a coherent field such that an effective 3-vertex appears.
To keep the discussion more general, we may write for the scaling behavior of a generic collision term for $l$-vertex scattering processes
\begin{equation}
C^{(l)}[f](t,{\mbf p}) \, = \, s^{-\mu_l}\, C^{(l)}[f](s^{-1/\beta}t,s \mbf p) \, ,
\end{equation}
where the scaling exponent
\begin{equation} 
\mu_l = (l-2) d - (l+1) - (l-1)\alpha/\beta\, 
\label{eq:mul}
\end{equation}
follows from similar arguments as exemplified for the \mbox{4-vertex} interaction.

Using the scaling relation (\ref{eq:scalingrelab}) and particle conservation (\ref{eq:pcrel}) gives the perturbative solution for 
\begin{equation} 
\mbox{\it rel.\ particle transport:}\,\,\, \alpha = - \frac{d}{l+1},\, 
\beta = -\frac{1}{l+1} .
\label{eq:nab}
\end{equation}
Similarly, with (\ref{eq:ecrel}) for relativistic $z=1$ one finds the perturbative solution for 
\begin{equation} 
\mbox{\it rel.\ energy transport:}\,\,\, \alpha = - \frac{d+1}{2l-1},\, \beta = -\frac{1}{2l-1} .
\label{eq:eab}
\end{equation}
For instance, for quartic self-interactions the perturbative energy transport is characterized by $\alpha = -(d+1)/7$, and $\beta=-1/7$, where the latter is independent of the dimensionality of space $d$. Likewise, for an effective 3-vertex in the presence of a coherent field one has for the energy transport $\alpha = -(d+1)/5$ and $\beta=-1/5$. Indeed, the latter values for the scaling exponents describe well the energy transport at higher momenta of the relativistic scalar field theory for $d=3$ of section~\ref{sec:rel}~\cite{Micha:2004bv}. In particular, their negative values indicate the direction of the transport from lower to higher momenta.\\

We now turn to the nonrelativistic theory of section~\ref{sec:nonrel}. The perturbative kinetic equation for $2 \leftrightarrow 2$ scattering is given again by (\ref{eq:2to2cl}), however, with the nonrelativistic 
\begin{eqnarray} 
&&\int {\mathrm{d}}\Omega_{\rm nr}^{2\leftrightarrow 2}({\mbf p},{\mbf l},{\mbf q},{\mbf r}) \sim g^2 \int\frac{d^d l}{(2\pi)^d}\frac{d^d q}{(2\pi)^d}\frac{d^d r}{(2\pi)^d}  \nonumber\\
&& \times(2\pi)^{d+1}\, \delta^{(d)}({\mbf p}+{\mbf l}-{\mbf q}-{\mbf r})\, 
\delta(\omega_{\mbf p} + \omega_{\mbf l} - \omega_{\mbf q} - \omega_{\mbf r})  \quad
\label{eq:nrmeasurepert}
\end{eqnarray}
in the absence of a condensate for $\omega_{\mbf p} = {\mbf p}^2/(2m)$. The scaling analysis follows along the same lines as before, but without the inverse-frequency factors from the relativistically invariant measure appearing in (\ref{eq:measurepert}). Accordingly, generalizing again to $l$-vertex scatterings, one obtains for the scaling relation~(\ref{eq:scalingrelab}) in the nonrelativistic case: $(l-2) \alpha = \beta [ (l-2) d - 2] - 1 $\, .
This leads with (\ref{eq:pcrel}) and (\ref{eq:ecrel}) to the solutions for
\begin{equation} 
\mbox{\it nonrel.\ particle transport:}\,\,\,\, \alpha = - \frac{d}{2},\,\,\, 
\beta = -\frac{1}{2} ,
\label{eq:nrnab}
\end{equation}
and 
\begin{equation} 
\mbox{\it nonrel.\ energy transport:}\,\, \alpha = - \frac{d+2}{2(l-1)},\, \beta = -\frac{1}{2(l-1)} .
\label{eq:nreab}
\end{equation} 
For instance, for quartic self-interactions the perturbative particle transport would be described by $\alpha = -d/2$, and $\beta=-1/2$.

We emphasize that all the above perturbative estimates with negative values for $\alpha$ and $\beta$ cannot account for the inverse particle transport observed from the simulations for $d=3$ in sections~\ref{sec:simcl} and \ref{sec:simreldyn}. Of course, perturbation theory is not expected to be applicable to the overoccupied infrared modes and one has to employ an alternative description, which we consider next.

\subsection{Scaling behavior in the overoccupied regime}
\label{sec:npscaling}

Remarkably, the overoccupied regime can still be described in terms of a generalized kinetic theory by taking into account vertex corrections. For the $N$-component field theory, these corrections can be systematically computed from an expansion in the number of field components $N$ to next-to-leading order~\cite{Berges:2001fi,Aarts:2002dj,Berges:2010ez}. The NLO corrections take scattering events up to infinite order into account. This allows us to describe even strongly correlated regimes, where the typical mode occupancies (\ref{eq:overpopulation}) or (\ref{eq:reloo}) are inversely proportional to the diluteness or coupling parameter, respectively. 

\begin{figure}[t!]
\centering
  \vspace*{-0.8cm}
	\includegraphics[width=.45\textwidth]{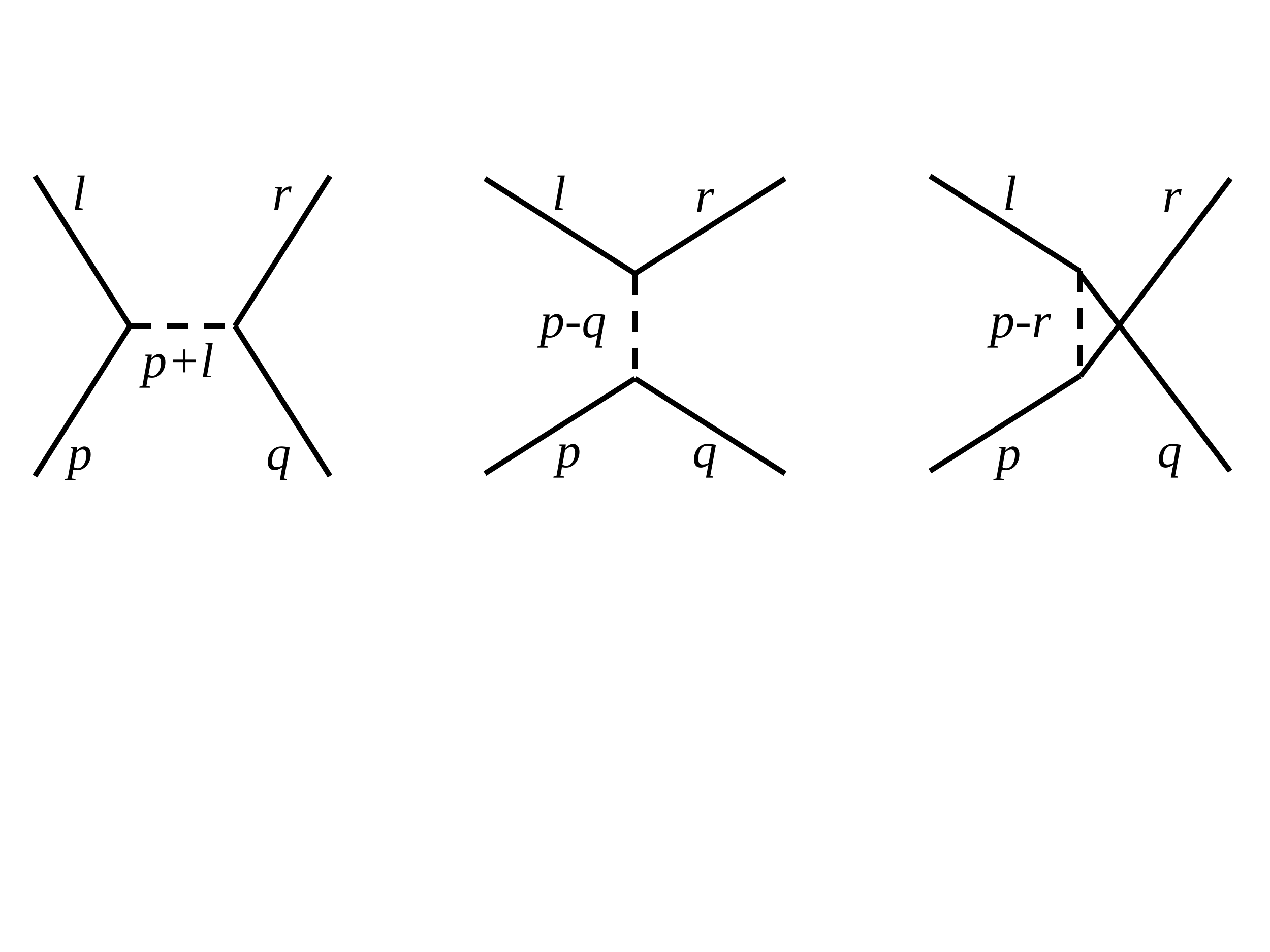}
	\vspace*{-2.6cm}
	\caption{Illustration of different scattering channels. The vertex correction at NLO may be viewed as an effective interaction, which involves the exchange of an intermediate particle. Left: The incoming particles with momenta $p$ and $l$ join into an intermediate particle that eventually splits. Middle and right: The incoming particle with momentum $p$ emits the intermediate particle and becomes the final particle with momentum $q$ or $r$, respectively.}
	\label{fig:channels}
\end{figure}

Effectively, the change to the perturbative kinetic equations (\ref{eq:2to2}) or (\ref{eq:2to2cl}) is the appearance of a time and momentum dependent matrix element squared:
\begin{equation} 
\lambda^2 \quad \rightarrow \quad \lambda^2_{\rm eff}[f](t,{\mbf p},{\mbf l},{\mbf q},{\mbf r}) \, .
\end{equation}
More precisely, one finds at NLO of the expansion in the number of field components for the relativistic theory a kinetic equation where (\ref{eq:measurepert}) is replaced by:
\begin{eqnarray} 
&& \int d\Omega^{\rm NLO}[f](t,{\mbf p},{\mbf l},{\mbf q},{\mbf r}) =  \frac{1}{6 N} \int\frac{d^d l}{(2\pi)^d}\frac{d^d q}{(2\pi)^d}\frac{d^d r}{(2\pi)^d}  \nonumber\\
&& \quad \times\, (2\pi)^{d+1}\, \delta^{(d)}({\mbf p}+{\mbf l}-{\mbf q}-{\mbf r})\, 
\frac{\delta(\omega_{\mbf p} + \omega_{\mbf l} - \omega_{\mbf q} - \omega_{\mbf r})}{2\omega_{\mbf p} 2\omega_{\mbf l} 2\omega_{\mbf q} 2\omega_{\mbf r}} \nonumber\\
&& \quad \times\, \lambda^2_{\rm eff}[f](t,{\mbf p},{\mbf l},{\mbf q},{\mbf r}) \, , 
\label{eq:measureNLO}
\end{eqnarray}
with dispersion $\omega_{\mbf p} = \sqrt{{\mbf p}^2 + m^2}$. Here 
\begin{eqnarray}
\lambda^2_{\rm eff}(t,{\mbf p},{\mbf l},{\mbf q},{\mbf r}) &\equiv& \frac{\lambda^2}{3}
\left[ \frac{1}{| 1 + \Pi^{R} (t,\omega_{\mbf p} + \omega_{\mbf l},{\mbf p}+{\mbf l})|^2} \right. \nonumber\\
&+& \frac{1}{| 1 + \Pi^{R} (t,\omega_{\mbf p} - \omega_{\mbf q},{\mbf p}-{\mbf q})|^2} 
\nonumber\\
&+&  \left. \frac{1}{| 1 + \Pi^{R} (t,\omega_{\mbf p} - \omega_{\mbf r},{\mbf p}-{\mbf r})|^2} \right] \quad
\label{eq:leffrel}
\end{eqnarray}
incorporates the vertex corrections for the different scattering channels. They are depicted in FIG.~\ref{fig:channels} and may be viewed as coming from an effective interaction, which involves the exchange of an intermediate particle whose four-momentum equals $p+l$, $p-q$ and $p-r$, respectively. The appearance of the `one-loop' retarded self-energy, 
\begin{eqnarray}
	&& \Pi^R(t,\omega,\mathbf p) \, = \, \frac{\lambda}{12} \int \frac{\mathrm{d}^dq}{(2\pi)^d}\, \frac{f(t,\mbf p-\mbf q)}{\omega_{\mbf q}\, \omega_{\mbf p-\mbf q}} \nonumber\\
	&& \quad \times \bigg[ \frac{1}{\omega_{\mbf q}+\omega_{\mbf p-\mbf q}-\omega} + \frac{1}{\omega_{\mbf q}-\omega_{\mbf p-\mbf q}-\omega}  \nonumber\\
	&& \quad \,\, +\frac{1}{\omega_{\mbf q}-\omega_{\mbf p-\mbf q}+\omega} + \frac{1}{\omega_{\mbf q}+\omega_{\mbf p-\mbf q}+\omega} \bigg] ,  \quad
\label{eq:pir_onshell_rel}
\end{eqnarray}
in the denominator is the result of a geometric series summation of the infinite number of NLO processes. It should be emphasized that $\Pi^R$, and thus also $\lambda^2_{\rm eff}$, is time-dependent because it is expressed in terms of the evolving distribution function. The above expressions correspond to the on-shell limit of the evolution equations presented in reference~\cite{Berges:2010ez} and their relation to the underlying field theory is further discussed in appendix~\ref{app:2pi}.

For sufficiently high momenta, the self-energy (\ref{eq:pir_onshell_rel}) becomes small such that
$\Pi^R(t,\omega,\mathbf p) \ll 1$ in the squared matrix element (\ref{eq:leffrel}) and we recover the perturbative expression $\lambda^2_{\rm eff} \rightarrow \lambda^2$. In this case, the kinetic equation corresponds to (\ref{eq:measurepert}) with the prefactor for large $N$. Since the scaling solutions of section~\ref{sec:perturbativescaling} do not depend on $N$, one gets the same results in the high momentum regime.
 
For low momenta the case $\Pi^R(t,\omega,\mathbf p) \gg 1$ can become relevant, which changes the situation dramatically. This becomes even more involved if a mass gap exists, as seen in the simulations of section~\ref{sec:massgap}. An effective mass gap is typically expected because of medium effects even if the mass parameter of the underlying microscopic theory is set to zero. In that case the infrared modes behave effectively nonrelativistically, which allows one to observe the same scaling behavior of the relativistic and the nonrelativistic theory in this regime. 

The corresponding kinetic equation for the nonrelativistic theory may be obtained from the relativistic case (\ref{eq:leffrel}), assuming that the mass appearing in the dispersion relation is much larger than the typical momenta. Expanding $\omega_{\mbf p} = \sqrt{{\mbf p}^2 + m^2} \simeq m + \mbf p^2/2m$ and inserting this into (\ref{eq:measureNLO}), we have $\delta(\omega_{\mbf p}+\omega_{\mbf l}-\omega_{\mbf q}-\omega_{\mbf r})\rightarrow \delta((\mbf p^2+\mbf l^2-\mbf q^2-\mbf r^2)/(2m))$
and $2\omega_{\mbf p} 2\omega_{\mbf l} 2\omega_{\mbf q} 2\omega_{\mbf r} \rightarrow 16 m^4$
to lowest nonvanishing order. Then $g \sim \lambda/m^2$ leads to the nonrelativistic form of the equation. Alternatively, one can obtain the corresponding result from a nonrelativistic $N$-component complex scalar field theory~\cite{Scheppach:2009wu} by performing again the $1/N$ expansion to NLO. This is further described in section~\ref{sec:anomalousscaling}.

Accordingly, we consider for the nonrelativistic case a kinetic equation with 
\begin{eqnarray} 
&& \int d\Omega_{\rm nr}^{\rm NLO}[f](t,{\mbf p},{\mbf l},{\mbf q},{\mbf r}) \sim \int\frac{d^d l}{(2\pi)^d}\frac{d^d q}{(2\pi)^d}\frac{d^d r}{(2\pi)^d}  \nonumber\\
&& \quad \times\, (2\pi)^{d+1}\, \delta^{(d)}({\mbf p}+{\mbf l}-{\mbf q}-{\mbf r})\, 
\delta(\omega_{\mbf p} + \omega_{\mbf l} - \omega_{\mbf q} - \omega_{\mbf r})  \nonumber\\[0.1cm]
&& \quad \times\,  g^2_{\rm eff}[f](t,{\mbf p},{\mbf q}) \, , 
\label{eq:measureNLOnr}
\end{eqnarray}
where $\omega_{\mbf p} = {\mbf p}^2/(2m)$ and
\begin{equation}
 g^2_{\rm eff}(t,{\mbf p},{\mbf q}) \equiv \frac{g^2}{| 1 + \Pi^{R}_{\rm nr} (t,\omega_{\mbf p} - \omega_{\mbf q},{\mbf p}-{\mbf q})|^2}
\label{eq:geff}
\end{equation}
incorporates the vertex corrections. The compact form of (\ref{eq:geff}) is achieved by using the symmetries of (\ref{eq:measureNLOnr}) and the nonrelativistic `one-loop' retarded self-energy reads 
\begin{eqnarray}
	&& \Pi^R_{\rm nr}(t,\omega, \mbf p) \,=\, 2g \int \frac{\mathrm{d}^dq}{(2\pi)^d}\, f(t,\mbf p-\mbf q) \nonumber\\
	&& \quad \times \bigg[ \frac{1}{\omega_{\mbf q}-\omega_{\mbf p-\mbf q}-\omega} + \frac{1}{\omega_{\mbf q}-\omega_{\mbf p-\mbf q}+\omega} \bigg] .
\label{eq:pir_onshell_nonrel1}
\end{eqnarray}
By comparison to its relativistic counterpart, this may also be obtained directly from (\ref{eq:pir_onshell_rel}) if evaluated as in (\ref{eq:geff}) by expanding the relativistic dispersion for small momenta and taking the dominant contributions, where the constant mass term cancels in the respective frequency sums.  

From the self-similar behavior of the distribution (\ref{eq:gen_selfsim_ansatz}), we can deduce the scaling property
\begin{equation}
\Pi^R_{\rm nr}(t,\omega_{\mbf p}, \mbf p) = s^{\alpha/\beta-d+2}\, \Pi^R_{\rm nr}(s^{-1/\beta} t,\omega_{s \mbf p}, s \mbf p) \, .
\end{equation}
Since for the relevant cases (\ref{eq:pcrel}) and (\ref{eq:ecrel}) we have \mbox{$\alpha/\beta \geq d$}, we conclude by keeping $s \mbf p$ fixed that $\Pi^R_{\rm nr}$ can become large in the infrared. In this case, we can use $\Pi^R_{\rm nr}(t,\omega_{\mbf p}, \mbf p) \gg 1$ to find the scaling behavior of
\begin{equation}
g^2_{\rm eff}(t,{\mbf p},{\mbf q},{\mbf r}) = s^{-2(\alpha/\beta-d+2)}\, g^2_{\rm eff}(s^{-1/\beta} t,s {\mbf p},s {\mbf q},s {\mbf r}) \, .
\end{equation}
Therefore, we expect the effective matrix element squared to become small in the infrared in accordance with related studies~\cite{Berges:2008wm}. In turn, we will see in the following that at the same time the distribution function $f(t,\mbf p)$ can grow significantly, which results in a `Bose enhancement' of scatterings that counteracts the diminished effective coupling.    

We use (\ref{eq:measureNLOnr}) in the corresponding kinetic equation (\ref{eq:2to2cl}) and find for the scaling of the collision term 
\begin{eqnarray}
C_{\rm nr}^{\rm NLO}[f](t,{\mbf p}) & = & s^{-(2-\alpha/\beta)}\, C_{\rm nr}^{\rm NLO}[f](s^{-1/\beta}t,s \mbf p) \nonumber\\
& = & t^{\alpha-2 \beta} \, C_{\rm nr}^{\rm NLO}[f_S](1,t^\beta \mbf p).
\end{eqnarray}
Remarkably, with this the exponent $\beta$ can be obtained solely from the scaling relation (\ref{eq:scalingrelab}) without using in addition energy or particle conservation, whereas the different solutions for $\alpha$ arise from imposing (\ref{eq:pcrel}) or (\ref{eq:ecrel}), respectively. We thus find in the overoccupied infrared regime for
\begin{equation} 
\mbox{\it nonrel.\ transport:}\,\,\, \beta =  \frac{1}{2} \,\,\, \mbox{\it of} \,
\left\{\! \begin{array}{ll}
\mbox{\it particles:} & \alpha = d/2\\[0.1cm]
\mbox{\it energy:} & \alpha = (d+2)/2
\end{array} \right.
\label{eq:nonreltrans}
\end{equation}
This is a central analytic result of this work. In contrast to the previously known negative scaling exponents from perturbative estimates given in section~\ref{sec:perturbativescaling}, one observes that the positive values of $\alpha$ and $\beta$ obtained here describe an {\it inverse} particle transport with growing occupation number in the infrared. The quantitative agreement of the NLO estimates $\alpha = 3/2$ and $\beta = 1/2$ for $d=3$ with the full simulation results of section~\ref{sec:simreldyn} for the relativistic and section~\ref{sec:simcl} for the nonrelativistic theory is remarkable. Both the approximate analytic and the full simulation results within their numerical accuracy indicate no strong dependence on $N$.\\

For comparison, we finally also analyze the relativistic kinetic equation with (\ref{eq:measureNLO}) in the absence of any mass gap. To this end, we use $\omega_{\mbf p}=|\mbf p|$ and proceed as in the nonrelativistic case to find scaling relations for $\alpha$ and $\beta$. With
\begin{equation}
C^{\rm NLO}[f](t,{\mbf p}) = t^{\alpha-\beta} \, C^{\rm NLO}[f_S](1,t^\beta \mbf p)
\end{equation}
we find for the 
\begin{equation} 
\mbox{\it rel.\ transport:}\,\,\, \beta =  1 \,\,\, \mbox{\it of} \,
\left\{\! \begin{array}{ll}
\mbox{\it particles:} & \alpha = d\\[0.1cm]
\mbox{\it energy:} & \alpha = d+1
\end{array} \right.
\label{eq:nonreltransrel}
\end{equation}
These estimates indicate that the simulation results of section~\ref{sec:simreldyn} for the relativistic theory cannot be interpreted in terms of massless scaling in the infrared, which is explained by the presence of a mass gap in section~\ref{sec:massgap}.

\section{Anomalous scaling}
\label{sec:anomalousscaling}

The kinetic description of section~\ref{sec:vrkt} assumes canonical values for the dynamic exponent $z$, which describes the scaling of the dispersion (\ref{eq:zdisp}). For the relativistic theory without a mass gap, $\omega_{\mbf p}= |\mbf p|$ is employed. For the Gross-Pitaevskii theory, in the presence of a condensate, the approximate (Bogoliubov) dispersion is given by~\cite{Bogoliubov} 
\begin{equation}
\omega(\mbf p) = \sqrt{\frac{\mbf p^2}{2m} \left(\frac{\mbf p^2}{2m} + 2g|\psi_0|^2\right)}  \, . 
\label{eq:bogdis} 
\end{equation}
At larger momenta, or in the absence of a condensate, one recovers $\omega_{\mbf p}={\mbf p}^2/(2m)$, while for low momenta one has $\omega_{\mbf p}=c|\mbf p|$ with $c\equiv\sqrt{g|\psi_0|^2/m}$.  

From a field theoretic point of view, these dispersions with integer-valued $z$ are implemented using a canonical spectral function, such as the free-field form 
\begin{equation}
\tilde{\rho}_0(p^0,\mbf p) = 2\pi\, {\rm sgn} (p^0)\, \delta\left( (p^0)^2 - \omega_{\mbf p}^2 \right)
\label{eq:freerho}
\end{equation}
for the relativistic theory. In principle, nonperturbative scaling phenomena may involve an anomalous scaling exponent for the full spectral function $\tilde{\rho}(p^0,\mbf p)$ of the interacting theory. Using spatial isotropy we may write 
\begin{equation}
\tilde{\rho}(p^0,\mbf p) \, = \, s^{2-\eta}\, \tilde{\rho}(s^z p^0,s \mbf p) \, , 
\label{eq:scalingrho}
\end{equation}
with a nonequilibrium `anomalous dimension' $\eta$. The dynamic scaling exponent $z$ appears since only spatial momenta are related by rotational symmetry and frequencies may scale differently also in the relativistic theory because of medium effects. 

Small discrepancies between the results of our full numerical simulations and the analytic estimates of previous sections could possibly be rooted in the canonical assumption of an integer-valued $z$ and $\eta = 0$. This concerns mainly the infrared regime, where strongly nonlinear behavior occurs.
For instance, corresponding infrared scaling phenomena in thermal equilibrium near continuous phase transitions can exhibit non-trivial scaling exponents with a noncanonical $z$ and a nonzero (but typically small) value for $\eta$, which is also captured by the NLO approximation employed~\cite{Alford:2004jj}. 

Therefore, we consider in this section a field theoretical calculation of the self-similar behavior near nonthermal fixed points taking into account the possibility of anomalous scaling. It is again based on the two-particle irreducible (2PI) generating functional in quantum field theory, which is expanded up to next-to-leading order in the number of field components $N$~\cite{Berges:2001fi,Aarts:2002dj}. However, without using the additional assumption of a canonical form for the spectral function underlying the kinetic theory of section~\ref{sec:vrkt}.

\subsection{Nonrelativistic field theory}

For the nonrelativistic scalar field theory we consider a single complex scalar field $\psi(x)$ with the action
\begin{align}
S[\psi,\psi^*]=&\, \int d^{d+1}x \left\{\psi^* \left( i\partial_{x^0} + \frac{\nabla^2}{2m} \right) \psi - \frac{g}{2}(\psi\psi^*)^2 \right\}.
\label{eq:nonrel_action}
\end{align}
In the classical approximation this gives rise to the Gross-Pitaevskii equation (\ref{eq:gpe}). In the corresponding quantum theory, the spectral function is given by the expectation value of the {\it commutator} of two Heisenberg field operators as
\begin{equation}
\tilde{\rho}_{ab}(x,y) = \langle[\hat{\psi}_a(x),\hat{\psi}_a^\dagger(y)]\rangle \, .
\label{eq:spectral_fct}
\end{equation}
Here the index notation for $a,b=1,2$ employs  
\begin{equation}
\hat{\psi}_1 \equiv \hat{\psi} \quad , \quad \hat{\psi}_2 \equiv \hat{\psi}^\dagger 
\label{eq:indexnotation}
\end{equation}
in order to have a compact notation for the four different two-point functions that can be built from two complex fields. 

The other set of linearly independent two-point functions can be conveniently expressed in terms of the expectation value of the {\it anticommutator} of two fields as 
\begin{equation}
	F_{ab}(x,y)=\frac{1}{2}\langle\{\hat{\psi}_a(x),\hat{\psi}_b^\dagger(y)\}\rangle_c,
\label{eq:stat_fct}
\end{equation}
where the subscript `c' refers to the connected correlator. While in thermal equilibrium the anticommutator and commutator expectation values are related by the fluctuation-dissipation relation, in general they are linearly independent for systems out of equilibrium.\footnote{For an introductory text, see e.g.\ reference~\cite{Berges:2004yj}.} The absence of a fluctuation-dissipation relation is a crucial property of the scaling behavior near the nonthermal fixed points discussed.  

The evolution equations for $F_{ab}(x,y)$ and $\tilde{\rho}_{ab}(x,y)$ at NLO in the 2PI $1/N$ expansion are known~\cite{Berges:2001fi,Aarts:2002dj,Scheppach:2009wu} and given in appendix~\ref{app:kadanoff} for completeness, where we also outline the employed gradient expansion to lowest order.  It is important to notice that, because of the lowest order gradient expansion, the spectral function is time-independent according to (\ref{eq:transport_rho_nonrel}). For the spatially homogeneous system, it is convenient to Fourier transform the two-point functions with respect to their relative coordinates $x-y$ and to define the `time' variable as $t = (x^0 + y^0)/2$. One obtains, using $(F_p)_{ab}\equiv F_{ab}(t,p)$ and $(\tilde{\rho}_p)_{ab}\equiv\tilde{\rho}_{ab}(p)$ as a compact matrix notation in $(a,b)$-index space:  
\begin{align}
	\frac{\partial }{\partial t}\text{Tr}\big[F_p\big]=&\, - \frac{2}{(2\pi)^{2d+2}} \int d^{d+1}q\,d^{d+1}l\,d^{d+1}r \nonumber\\
	&\, \times \delta^{(d+1)}(p-q+l-r)\,\, g^2_{\text{eff}}[F](t,p-q) \nonumber\\
	&\, \times \bigg\{ \,2\,\text{Tr}\big[ \sigma^3F_p\,F_q \big] \text{Tr}\big[ F_l\,\tilde{\rho}_r \big] \nonumber\\
	&\ \  +\text{Tr}\big[ \sigma^3F_p\,\tilde{\rho}_q \big]\text{Tr}\big[ F_l\,F_r \big] \nonumber\\
	&\ \ -\text{Tr}\big[ \sigma^3\tilde{\rho}_p\,F_q \big] \text{Tr}\big[ F_l\,F_r \big] \bigg\},
\label{eq:kineq_offshell_nonrel}
\end{align}
where $\text{Tr}[F_p]\equiv F_{aa}(t,p)$ and $\sigma^3 = {\rm diag}(1,-1)$ denotes the third Pauli matrix. The time- and momentum-dependent effective coupling squared corresponding to (\ref{eq:geff}) reads
\begin{equation}
	g^2_{\text{eff}}(t,p)=\frac{g^2}{|1+\Pi^R_{\rm nr}(t,p)|^2}
\label{eq:effective_coupling1}
\end{equation}
with the retarded self-energy
\begin{equation}
	\Pi^R(t,p)= 2g \int \frac{d^{d+1}q}{(2\pi)^{d+1}} F_{ab}(t,q-p)\, G^R_{ba}(q) 
\label{eq:pir_nonrel1}
\end{equation}
in terms of the retarded propagator $G^R_{ab}(p)$ as is further discussed in appendix~\ref{app:2pi}.

To make contact with the definition of the distribution function in (\ref{eq:n_def_stat_nonrel}), we note that with the notation $F \equiv F_{aa}/2$ this can be written for ${\mbf p} \neq 0$ as:
\begin{equation}
	f(t,\mbf p) +\frac{1}{2} =  \frac{1}{2}\int \frac{dp^0}{2\pi} F_{aa}(t,p^0,\mbf p) ,
\label{eq:effpart_nonrel}
\end{equation}
where we write the `quantum-half' for completeness though we always consider high typical occupancies such that it can be neglected. With the field theoretical definition of the distribution function at hand, we can obtain a corresponding `collision term' for $\partial f(t,\mbf p)/\partial t = C[F](t,\mbf p)$ from the $p^0$-integral of the RHS of (\ref{eq:kineq_offshell_nonrel}). 

Applying (\ref{eq:gen_selfsim_ansatz}) to the distribution function with $f(t,\mbf p) \gg 1$ for typical momenta, self-similar behavior for the correlators is described by 
\begin{align}
	F_{ab}(t,p^0,\mbf p)&=s^{\alpha/\beta+z}\,F_{ab}(s^{-1/\beta}t,s^zp^0,s\mbf p) \, 
\label{eq:ansatz_F_nonrel}
\end{align}
in addition to the scaling behavior (\ref{eq:scalingrho}) for the spectral function $\tilde{\rho}_{ab}(p^0,\mbf p)$. This means $F_{ab}(t,p^0,\mbf p)=t^{\alpha+\beta z}\, F_{S,ab}(t^{z\beta}p^0,t^\beta\mbf p)$ with $F_{S,ab}(p^0,\mbf p)\equiv F_{ab}(1,p^0,\mbf p)$. 
In particular, $f_S(\mbf p)=\int dp^0/(2\pi) F_{S,aa}(p^0,\mbf p)/2$ in this highly occupied scaling regime according to (\ref{eq:effpart_nonrel}). With
\begin{equation}
	g^2_{\text{eff}}[F](t,p^0,\mbf p)=t^{-2[\alpha+\beta(2-\eta-d)]}\,g^2_{\text{eff}}[F_S](t^{z\beta} p^0,t^\beta\mbf p)
\label{eq:scaling_leff_offshell_nonrel}
\end{equation}
one finds
\begin{align}
	C[F](t,\mbf p)=&\,t^{\alpha-\beta(2-\eta)}\,C[F_S](t^\beta\mbf p).
\label{eq:rhs_np_offshell}
\end{align}

We can now proceed in complete analogy to the analysis of section~\ref{sec:vrkt} from which we find the solution to the scaling relation:
\begin{align}
	\beta=&\,\frac{1}{2-\eta} \, ,
\label{eq:beta_np_nonrel}
\end{align}
where the dimensionality $d$ and the exponent $\alpha$ have dropped out. 
In addition we have the time-independent equation for the nonthermal fixed point function:
\begin{align}
	\big[\alpha+\beta\,\mbf p\cdot\nabla_{\mbf p}\big]f_S(\mbf p)=&\,C[F_S](\mbf p)\, 
\label{eq:unifunc_offshell_nonrel}
\end{align}
corresponding to (\ref{eq:fpequation}).

To obtain a second scaling relation for the determination of $\alpha$, it is important to note that the particle number density $n = \int d^{d+1}p/(2\pi)^{d+1} F_{aa}(t,p)/2$ and the energy density $\epsilon =  \int d^{d+1}p/(2\pi)^{d+1} p^0 F_{aa}(t,p)/2$ are conserved during the evolution. We find: 
\begin{align}
	\alpha=&\,\frac{d}{2-\eta}  &\text{\it (particle transport)},
\label{eq:alpha_part_nonrel}\\
	\alpha=&\,\frac{d+z}{2-\eta} &\text{\it (energy transport)}.
\label{eq:alpha_energy_nonrel}
\end{align}
Of course, taking $\eta=0$ and $z=2$ we recover the results of section~\ref{sec:npscaling}.
Remarkably, one observes that the scaling exponent for the nonrelativistic particle cascade is independent of the dynamic exponent $z$. This has the important special consequence that the same scaling behavior for particle transport is found no matter whether the low-momentum dispersion is quadratic, or linear as for the Bogoliubov dispersion (\ref{eq:bogdis}) in the presence of a condensate. 

Comparing the expressions for particle transport to our simulation results of section~(\ref{sec:simcl}), within the numerical errors we see no strong indication for any deviation from the canonical value $\eta=0$ employed for the analytical estimates of section~\ref{sec:vrkt}. 

We may also consider the growth of the condensate during the self-similar evolution, which we  investigated numerically in sections~\ref{sec:nonrel} and \ref{sec:rel}. With $F_{ab}(t,t,\mbf{p})\sim\int dp^0\,F_{ab}(t,p^0,\mbf{p})$ and using (\ref{eq:ansatz_F_nonrel}) we obtain the scaling of a characteristic mode of the equal-time correlator $\sim t^{\alpha}$.
Since $\alpha>0$ for not too large $\eta$, the condensate zero-mode is expected to grow as a power-law in time. The value $\alpha \simeq 3/2$ for $d=3$ rather accurately describes our numerical findings.

For completeness, we indicate how to recover the kinetic equation of section~\ref{sec:npscaling} from the evolution equation (\ref{eq:kineq_offshell_nonrel}) for the anticommutator expectation value. First, one inserts the free-field form of the spectral function~\cite{Branschadel:2008sk}
\begin{align}
	\tilde{\rho}(p)=&\, 2\pi \left( \begin{matrix} \delta(p^0-\omega_{\mbf p}) & 0 \\ 0 & -\delta(p^0+\omega_{\mbf p}) \end{matrix} \right),
\label{eq:free_rho_nonrel_high}
\end{align}
where $\omega_{\mbf p}=\mbf p^2/2m$. Second, one can define an off-shell distribution function $f(t,p)$ as
\begin{align}
	F_{ab}(t,p)=&\, \left(f(t,p)+\frac{1}{2}\right)\, \tilde{\rho}_{ab}(p) \, 
\label{eq:offshell_partnumb_def}
\end{align}
with $-f(t,-p)=f(t,p)+1$ in accordance to (\ref{eq:effpart_nonrel}).
Inserting all this yields for $f(t,\mbf p) \gg 1$ the nonrelativistic kinetic equation employed in section~\ref{sec:npscaling}.

\subsection{Relativistic field theory}
\label{sec:relft}

We now consider the relativistic quantum field theory for a $N$-component real scalar field $\varphi_a(x)$ with $a\in\{1,\ldots,N\}$ in $d$ spatial dimensions. Its classical action is given by
\begin{align}
S=&\int d^{d+1}x \left\{ \frac{1}{2}\partial^\mu\varphi_a\partial_\mu\varphi_a - \frac{m^2}{2}\varphi_a\varphi_a - \frac{\lambda}{4!N}\big( \varphi_a\varphi_a  \big)^2 \right\}.
\label{eq:rel_action}
\end{align}
The corresponding classical equation of motion for $m=0$ is the Klein-Gordon equation given in (\ref{eq:kge}). 

Following the discussion of section~\ref{sec:npscaling} we will focus here on the case without a mass gap, since otherwise the above nonrelativistic results apply. Along the lines of the previous section, we introduce spectral and statistical functions as the commutator and connected anticommutator expectation values of two Heisenberg field operators. In the following we will assume that the corresponding $F_{ab}=F\,\delta_{ab}$ and $\tilde{\rho}_{ab}=\tilde{\rho}\,\delta_{ab}$. According to (\ref{eq:reldeffcond}) and taking into account the quantum-half, we get the distribution function for sufficiently high occupancies for ${\mbf p} \neq 0$ as:
\begin{align}
	f(t,\mbf p) + \frac{1}{2} =&\, \int_0^\infty\frac{dp^0}{2\pi}2p^0F(t,p) \, .
\label{eq:effpart_rel}
\end{align}

The derivation of the evolution equation for $f(t,\mbf p)$ follows along the same lines as for the nonrelativistic theory of the previous section. Using the 2PI $1/N$ expansion to NLO, the lowest-order gradient expansion leads to
\begin{eqnarray}
&&	\frac{\partial f}{\partial t}(t,\mbf p) \,=\, \frac{1}{6N(2\pi)^{2d+3}} \int_0^\infty dp^0dl^0dq^0dr^0 \int d^dl\, d^dq\, d^dr \nonumber\\
&& \quad\times\, \delta^{(d+1)}(p+l-q-r)\, \lambda^2_{\text{eff}}[F](t,p,l,q,r)\nonumber\\[3pt]
&&\quad \times \Big[ \big( \tilde{\rho}_p\,F_l+F_p\,\tilde{\rho}_l \big)F_qF_r-F_pF_l \big( \tilde{\rho}_q\,F_r+F_q\,\tilde{\rho}_r \big) \Big],
\label{eq:kineq_offshell_rel}
\end{eqnarray}
where $F_p=F(t,p)$, and $\lambda^2_{\text{eff}}$ is defined as
\begin{eqnarray}
\lambda^2_{\mathrm{eff}}(t,p,l,q,r) &=& \frac{\lambda^2}{3} \left[ {v_{\mathrm{eff}}(t,p+l)} \right.
\nonumber\\
&+& \left. {v_{\mathrm{eff}}(t,p-q)} + {v_{\mathrm{eff}}(t,p-r)} \right]  
\label{eq:lamdaeffdef}
\end{eqnarray}
with the vertex function $v_{\mathrm{eff}}(t,p)$ given in (\ref{eq:effective_coupling}).

With the above kinetic equation we now search for self-similar solutions (\ref{eq:gen_selfsim_ansatz}). For this we have to write
\begin{align}
	F(t,p^0,\mbf p)&=s^{\alpha/\beta+2z}\,F(s^{-1/\beta}t,s^zp^0,s\mbf p),
\label{eq:ansatz_F_rel}
\end{align}
in addition to the scaling behavior (\ref{eq:scalingrho}) for the spectral function.
Inserting the scaling behavior into the evolution equation (\ref{eq:kineq_offshell_rel}) leads to the solution
\begin{align}
	\beta=&\,\frac{1}{2-\eta-z},
\label{eq:beta_np_rel}
\end{align}
and an equation for the universal scaling function corresponding to (\ref{eq:unifunc_offshell_nonrel}). Imposing particle number and energy conservation yields
\begin{align}
	\alpha=&\,\frac{d}{2-\eta-z} &\text{\it (particle transport)},
\label{eq:alpha_part_rel}\\
	\alpha=&\,\frac{d+z}{2-\eta-z} &\text{\it (energy transport)}.
\label{eq:alpha_energy_rel}
\end{align}
For $\eta=0$ and $z=1$ these values agree with the results in the absence of a mass gap of section~\ref{sec:npscaling}. We finally note that the relativistic kinetic equation of that section can be obtained from (\ref{eq:kineq_offshell_rel}) by using the corresponding definition (\ref{eq:offshell_partnumb_def}) with the free spectral function (\ref{eq:freerho}).

\section{Conclusion}

A consistent picture has emerged for the universal self-similar dynamics of relativistic and nonrelativistic field theories near nonthermal fixed points. The results of full numerical simulations are well described by analytic results based on the vertex-resummed kinetic theory. The latter extends well-established kinetic descriptions to the nonperturbative regime of overoccupied modes. 

The vertex-resummed kinetic theory links the perturbative phenomenon of weak wave turbulence relevant at higher momenta to the nonperturbative physics of strong turbulence and vorticity of the underlying field configurations in the highly nonlinear infrared regime. It is striking that the range of validity of kinetic descriptions can indeed be extended to capture these very different regimes.

For the examples of nonrelativistic (Gross-Pitaevskii) and relativistic scalar field theory with quartic self-interactions, we have seen that the infrared scaling exponents as well as scaling functions agree. This becomes possible because of the emergence of a mass gap in the relativistic theory. In contrast to the previously known negative values for the scaling exponents $\alpha$ and $\beta$ from perturbative estimates, we find that their positive values $\alpha = d/(2-\eta)$ and $\beta = 1/(2-\eta)$ obtained for small anomalous dimension $\eta$ describe an inverse particle transport with growing occupation number in the infrared. The growth exponent $\alpha$ is found to describe also the far-from-equilibrium formation of the Bose condensate. 

Moreover, this nonrelativistic particle transport solution has the remarkable property to be independent of the dynamic scaling exponent $z$. As a consequence, this solution applies equally well to a dispersion with quadratic momentum dependence or a possible linear behavior below the characteristic coherence momentum scale in the presence of a Bose gas condensate.

The corresponding dynamic universality class turns out to be remarkably large, encompassing both relativistic as well as nonrelativistic quantum and classical systems. As a consequence, the applications can range from table-top experiments with ultracold quantum gases to inflationary dynamics during the very early stages of our universe.

\begin{acknowledgments}
We thank I.~Chantesana, T.~Gasenzer, V.~Kasper, S.~Schlichting, D.~Sexty, R.~Venugopalan and B.~Wallisch for discussions and collaborations on related work. This work was supported in part by the German Research Foundation (DFG). K.~B. thanks HGS-HIRe for FAIR for support. Part of the numerical results presented in this work were obtained on the bwGRiD (\url{http://www.bw-grid.de}), member of the German D-Grid initiative, funded by the Ministry for Education and Research (BMBF) and the Ministry for Science, Research and Arts Baden-W\"urttemberg (MWK-BW).
\end{acknowledgments}

\appendix

\section{Stationary transport}
\label{eq:waveturbulence}

For comparison, we summarize in this section the analysis for stationary transport. This is a time-translation invariant problem, which can be realized in the presence of suitable sources and sinks. In contrast, the isolated systems out of equilibrium considered in the main part of this work are characterized by a self-similar evolution, which is not time-translation invariant. 

Stationary transport in perturbative kinetic theory has been intensively studied in the context of weak wave turbulence~\cite{Zakharov:1992,Nazarenko:2011,Micha:2004bv}.
Similarly, the nonperturbative regime of strong turbulence has been investigated in detail and the following discussion is based on the literature~\cite{Berges:2008wm,Berges:2008sr,Scheppach:2009wu}, where we point out the relevant aspects for the comparison to the case of self-similar evolutions.

\subsection{Perturbative regime: Weak wave turbulence}
\label{eq:weakwaveturbulence}

In perturbative kinetic theory, when two particles scatter into two particles, the time evolution of the distribution function $f_{\mbf p}(t)$ for a spatially homogeneous system is given for not too small occupation numbers by (\ref{eq:2to2cl}). For the example of the relativistic $N$-component scalar field theory with quartic $\lambda/(4!N)$-interaction this is further specified by (\ref{eq:measurepert}). The phenomenon of weak wave turbulence is expected for occupation numbers per mode in the regime $1 \ll f_{\mbf p} \ll 1/\lambda$, where that perturbative description is expected to be valid. In section~\ref{sec:nonpert} below, we consider the nonperturbative case of an overoccupied system. 

For number conserving $2 \leftrightarrow 2$ scatterings, apart from the energy density $\epsilon$ also the total particle number density $n$ is conserved, which are given by
$\epsilon = \int_{\mbf p} \omega_{\mbf p} f_{\mbf p}$ and $n = \int_{\mbf p} f_{\mbf p}$ with the notation $\int_{\mbf p} \equiv \int d^d p/(2\pi)^d$. The fact that they are conserved 
may be described by a continuity equation in momentum space, such as 
\begin{equation} 
\frac{\partial}{\partial t}\left( \omega_{\mbf p} f_{\mbf p} \right) + \nabla_{\mbf p} \cdot {\bf j}_{\mbf p} \, = \, 0
\label{eq:continuity}
\end{equation}
for energy conservation. Similarly, particle number conservation is described by formally replacing $\omega_{\mbf p} \to 1$ in the above equation and a corresponding 
substitution of the flux density. For the isotropic situation we can consider the energy flux $A(k)$ through a momentum sphere of radius $k$. Then only the radial component of the flux density ${\mbf j}_{\mbf p}$ is nonvanishing and
\begin{equation} 
\int_{\mbf p}^k \nabla_{\mbf p} \cdot {\mbf j}_{\mbf p} \, = \, \int_{\partial k} {\mbf j}_{\mbf p} \cdot {\mathrm d}{\mathbf{A}}_{\mbf p} \, \equiv \, (2\pi)^d A(k)\, .
\end{equation}
Since in this approximation $\omega_{\mbf p}$ is constant in time, we can write with the help of (\ref{eq:continuity}) and the kinetic equation: 
\begin{equation} 
A(k) \, = \, - \frac{1}{2^d \pi^{d/2} \Gamma(d/2+1)}\, \int^k \! {\mathrm d}p \, |{\mbf p}|^{d-1} \omega_{\mbf p} C^{2\leftrightarrow 2}({\mbf p}) \, .
\label{eq:flux}
\end{equation}
Stationary wave turbulence is characterized by a {\it scale-independent flux} $A(k)$, for which the respective integral does not depend on the integration limit $k$. To this end, one considers scaling solutions 
\begin{equation} 
f_{{\mbf p}} \, = \, s^{\kappa} f_{s{\mbf p}} \quad , \quad 
\omega_{{\mbf p}} \, = \, s^{-1}\, \omega_{s{\mbf p}} \, ,
\label{eq:scalingprop}
\end{equation}
with occupation number exponent $\kappa$ and assuming a linear dispersion relation relevant for momenta $|{\mbf p}| \gg m$.
Since the physics is scale invariant, we can choose $s = 1/|{\mbf p}|$ such that $f_{\mbf p} = |{\mbf p}|^{-\kappa}\, f_1$ and $\omega_{\mbf p} = |{\mbf p}|\, \omega_1$.

Using these scaling properties, one obtains for the collision integral (\ref{eq:measurepert}) and (\ref{eq:2to2cl}) of the theory with quartic self-interaction:
\begin{equation} 
C^{2\leftrightarrow 2}({\mbf p}) \, = \, s^{-\tilde{\mu}_4}\, C^{2\leftrightarrow 2}(s{\mbf p}) \, , 
\end{equation}
where the scaling exponent is given by 
\begin{equation} 
\tilde{\mu}_4 \, = \, (3d-4)-(d+1) - 3\kappa\, = \, 2d - 5 - 3\kappa\, .
\label{eq:tildeDel4}
\end{equation}
The first term in brackets comes from the scaling of the integral measure, the second from energy-momentum conservation for two-to-two scattering and the third from the three factors of the distribution function appearing in (\ref{eq:2to2cl}). 

Similar to the discussion in section~\ref{sec:perturbativescaling}, to be more general we may write for the scaling behavior of a generic collision term for $l$-vertex scattering processes
\begin{equation}
 C({\mbf p}) = |{\mbf p}|^{\tilde{\mu}_l}\, C({\mathbf 1}) 
\end{equation}
in terms of the scaling exponent $\tilde{\mu}_l$. One obtains along these
lines
\begin{equation} 
\tilde{\mu}_l = (l-2) d - (l+1) - (l-1) \kappa\, .
\label{eq:tildemul}
\end{equation}
For the scaling properties of the energy flux we can then write
\begin{eqnarray} 
A(k) & = & - \frac{1}{2^d \pi^{d/2} \Gamma(d/2+1)}\, \int^k \! {\mathrm d}p\, |{\mbf p}|^{d+\tilde{\mu}_l} \, \omega_1  C({\mathbf 1}) . \quad 
\label{eq:intA}
\end{eqnarray}
If the exponent in the integrand is nonvanishing, the integral gives
\begin{equation} 
A(k) \, \sim \,  \frac{k^{d+1+\tilde{\mu}_l}}{d+1+ \tilde{\mu}_l}\, \omega_1 C({\mathbf 1}) \, .
\end{equation} 
Thus, scale invariance may be obtained for
\begin{equation} 
d+1+\mu_l \, = \, 0 \, .
\end{equation}
This gives the scaling exponent for the perturbative 
\begin{equation} 
\mbox{\it relativistic energy cascade:} \,\,\, \kappa \, = \, d - \frac{l}{l-1} \, .
\label{eq:ekappaweak}
\end{equation}
One observes that stationary turbulence requires in this case the existence of the limit
\begin{equation} 
\lim_{d+1+\tilde{\mu}_l \to 0} \,\, \frac{C({\mathbf 1})}{d+1+\tilde{\mu}_l} \, = \, {\mathrm{const}} \, \neq \, 0 \, ,
\end{equation}
such that the collision integral must have a corresponding zero of
first degree. Similarly, starting from the continuity equation for
particle number one can study stationary turbulence associated to
particle number conservation. This leads to the perturbative 
\begin{equation} 
\mbox{\it relativistic particle cascade:}\,\,\, \kappa\, = \, d - \frac{l+1}{l-1} \, .
\end{equation}
Accordingly, for quartic self-interactions we get $\kappa = d - 4/3$ for the energy cascade and $\kappa = d - 5/3$ for the particle cascade. In the presence of a coherent field, when a 3-vertex can become relevant, the associated scaling exponents are $\kappa = d - 3/2$ for the energy cascade and $\kappa = d - 2$ for the particle cascade. For $d=3$ this result from stationary turbulence compares reasonably well with the approximate power law observed in section~\ref{sec:simreldyn} for the case of a self-similar evolution. For the corresponding perturbative nonrelativistic results we refer to the literature~\cite{Zakharov:1992,Nazarenko:2011,Micha:2004bv}.

\subsection{Nonperturbative regime: Strong turbulence}
\label{sec:nonpert}

The above perturbative description is expected to become invalid at sufficiently low momenta. In particular, the occupation numbers $f_{\mbf p} \sim |{\mbf p}|^{-\kappa}$ for $\kappa > 0$ would grow nonperturbatively large in the infrared such that the approximation (\ref{eq:2to2}) becomes questionable. To understand where the picture of weak wave turbulence breaks down and to compute the properties of the infrared regime, we have to consider nonperturbative approximations. For this purpose, we employ the vertex-resummed theory as in section~\ref{sec:anomalousscaling} based on the expansion of the 2PI effective action in the number of field components to NLO. 

For a relativistic theory with dispersion $\omega_{\mbf p} = \sqrt{{\mbf p}^2 + m^2}$, the scaling assumption (\ref{eq:scalingprop}) should be valid for sufficiently high momenta $|{\mbf p}| \gg m$ such that the dispersion is approximately linear with $\omega_{\mbf p} \sim |{\mbf p}|$. However, if a mass gap exists the infrared modes behave effectively nonrelativistic as explained in section~\ref{sec:npscaling}. In the following, we will analyze the two cases of a relativistic theory without mass gap and a nonrelativistic theory separately for comparison. 
 
We first look for relativistic scaling solutions in the absence of a mass gap. To be able to cope with occupancies of order $\sim 1/\lambda$, we replace the perturbative collision term by the vertex-resummed expression according to (\ref{eq:kineq_offshell_rel}) with 
\begin{eqnarray} 
&& \int {d}\Omega^{{\mathrm{NLO}}}(p,l,q,r) \, \simeq \, \frac{1}{6 N} \int_0^\infty \frac{{d}p^0{d}l^0{d}q^0{d}r^0}{(2\pi)^{4-(d+1)}} \nonumber\\ 
&& \times \int_{\mbf l \mbf q \mbf r} \delta(p+l-q-r) \tilde{\rho}_p \tilde{\rho}_l \tilde{\rho}_q \tilde{\rho}_r\, \lambda^2_{\mathrm{eff}}(p,l,q,r)\, ,
\label{eq:NLOomega}
\end{eqnarray}
where $\lambda^2_{\mathrm{eff}}$ is defined in (\ref{eq:lamdaeffdef}).
We emphasize that this still involves integration over frequencies as well as spatial momenta since no free-field form for the spectral function $\tilde{\rho}_p \equiv \tilde{\rho}(p^0,{\mbf p})$ is used so far, and $f_p \equiv f(p^0,{\mbf p})$. A crucial difference to the perturbative kinetic equation is the appearance of the vertex function $v_{\mathrm{eff}}(p^0,{\mbf p})$, which encodes the emergence of a momentum-dependent effective coupling from the NLO corrections of the $1/N$ expansion. 

We characterize the scaling properties of the spectral function $\tilde{\rho}(p^0,{\mbf p})$ as in (\ref{eq:scalingrho}), which takes into account a possible anomalous dimension $\eta$. Accordingly, the scaling behavior of the statistical correlation function
\begin{equation}
F(p^0,{\mbf p}) \, = \, s^{2+\kappa_{\mathrm s}}\, F(s^z p^0,s {\mbf p})
\label{eq:scalingF}  
\end{equation}
will be described using a scaling exponent $\kappa_{\mathrm s}$. This translates with the definition for $f_p \gg 1$ into
\begin{equation}
f(p^0,{\mbf p}) \, = \, s^{\kappa_{\mathrm s} + \eta}\, f(s^z p^0, s{\mbf p}) \, .
\label{eq:scalingn}
\end{equation}
Using these definitions, one can infer the scaling behavior of $v_{\mathrm{eff}}(p^0,{\mbf p})$, which is given in terms of the ``one-loop'' retarded self-energy $\Pi_{R}(p^0,{\mbf p})$ as in section~\ref{sec:relft} from which follows
\begin{equation}
\Pi_{R}(p^0,{\mbf p}) \, = \, s^\Delta\, \Pi_{R}(s^z p^0, s{\mbf p})
\end{equation}
with  
\begin{equation}
\Delta = 4 - d -z + \kappa_{\mathrm s} - \eta .
\label{eq:delta}
\end{equation}
If $\Delta > 0$ one finds the infrared scaling behavior
\begin{equation}
v_{\mathrm{eff}}(p^0,{\mbf p}) \, = \, s^{-2\Delta}\, v_{\mathrm{eff}}(s^z p^0, s{\mbf p}) \, . 
\end{equation}
(For $\Delta \leq 0$ the effective coupling would become trivial with $v_{\mathrm{eff}} \simeq 1$, on which we comment below.) 
Employing these scaling properties, (\ref{eq:NLOomega}) gives
\begin{eqnarray}
&& \int {d}\Omega^{{\mathrm{NLO}}}(p,l,q,r) \, = \, s^{-2\kappa_{\mathrm s}-z-2\eta} \nonumber\\
&& \times \int {d}\Omega^{{\mathrm{NLO}}}(s^z p^0,s^z l^0,s^z q^0,s^z r^0;s{\mbf p},s{\mbf l},s{\mbf q},s{\mbf r}). \quad
\end{eqnarray}

Following the procedure of section~\ref{eq:waveturbulence}, for any conserved quantity we can compute the flux through a momentum sphere $k$. Stationary turbulence solutions then require that the respective integral does not depend on $k$. Similar to (\ref{eq:flux}), the flux for this effective particle number reads 
\begin{equation}
A(k) \, = \, - \frac{1}{2^d \pi^{d/2} \Gamma(d/2+1)}\, \int^k \! {d}p\, |{\mbf p}|^{d-1}\, C^{{\mathrm{NLO}}}({\mbf p}) \, .
\label{eq:Aeff}
\end{equation}
The momentum integral can be evaluated along similar lines as before using the above scaling properties 
such that 
\begin{equation}
A(k) \, \sim \,  \frac{k^{d-\kappa_{\mathrm s}+z-\eta}}{d-\kappa_{\mathrm s}+z-\eta}\, C^{{\mathrm{NLO}}}({\mathbf 1}) \, .
\end{equation} 
Therefore, scale invariance may be obtained for the~\cite{Berges:2008wm,Berges:2008sr}
\begin{equation} 
\mbox{\it particle cascade:}\,\,\,  \kappa_{\mathrm s} \, = \, d + z - \eta  
\label{eq:kappasn}
\end{equation}
in the nonperturbative low-momentum regime.
Similarly, for the scaling solution associated to energy conservation one finds, taking into account the additional power of $p^0$ in the respective integrand, the exponent for the~\cite{Berges:2008wm,Berges:2008sr}
\begin{equation} 
\mbox{\it energy cascade:} \,\,\, \kappa_{\mathrm s} \, = \, d + 2z - \eta \, .
\label{eq:kappase}
\end{equation}
With these solutions, we can now reconsider the above assumption that $\Delta > 0$ by plugging (\ref{eq:kappasn}) or (\ref{eq:kappase}) into (\ref{eq:delta}). Indeed, it is fulfilled under the sufficient condition that $\eta < 2$. Taking into account that the anomalous dimension for scalar field theory is expected to be small for not too low dimension, we find $\kappa_{\mathrm s} \simeq d+1$ for the particle cascade and $\kappa_{\mathrm s} \simeq d +2$ for the energy cascade solution. 

The scaling of the effective occupation number distribution $f({\mbf p})$, which depends only on spatial momentum, can finally be obtained from 
\begin{eqnarray}
f({\mbf p}) + \frac{1}{2} &=& \int_0^\infty \frac{d p^0}{2\pi}\, 2  p^0 \tilde{\rho}(p^0,{\mbf p})\, \left[f(p^0,{\mbf p}) + \frac{1}{2}\right] \nonumber\\
&=&  \int_0^\infty \frac{{d} p^0}{2\pi}\, 2  p^0 F(p^0,{\mbf p}) , 
\label{eq:frel}
\end{eqnarray}
using that $\int_0^\infty d p^0/(2\pi) p^0 \tilde{\rho}(p^0,{\mbf p}) = 1/2$, which corresponds to the field commutation relation in Fourier space. Of course, the scaling behavior (\ref{eq:scalingF}) for $F(p^0,{\mbf p})$ may only be observed in a momentum regime with sufficiently high occupancies $f({\mbf p})$ as implied in the above derivation. Then (\ref{eq:frel}) yields:  
\begin{eqnarray}
f({\mbf p})  &=& s^{\kappa_{\mathrm s} + 2 - 2z} f(s{\mbf p}) \nonumber\\
&\sim&
\left\{ \begin{array}{ll}
|{\mbf p}|^{-(d+2-z-\eta)} & \mbox{\it rel.\ particle cascade}\\[0.1cm]
|{\mbf p}|^{-(d+2-\eta)} & \mbox{\it rel.\ energy cascade}
\end{array} \right.
\, 
\label{eq:relfscal}
\end{eqnarray}
These estimates show that vertex corrections can lead to a strongly modified infrared scaling behavior as compared to the perturbative treatment of section~\ref{eq:weakwaveturbulence}.   

We now turn to the nonrelativistic limit, which is outlined in section~\ref{sec:vrkt} and we present here the relevant changes as compared to the relativistic case. To this end, we consider a nonrelativistic $N$-component complex scalar field theory and perform again the $1/N$ expansion to NLO. Since already the relativistic scaling exponents (\ref{eq:relfscal}) indicate no explicit dependence on $N$ at NLO -- it can only enter indirectly via $\eta$ and $z$ -- this seems to be a very good starting point to understand also the single complex field case of Gross-Pitaevskii.   

Therefore, proceeding in the same way as for the relativistic theory, with the corresponding scaling ansatz for the spectral and statistical two-point functions, leads to the very same solutions (\ref{eq:kappasn}) and (\ref{eq:kappase}). A crucial difference arises when the occupation number distribution is determined. Using the nonrelativistic definition for the distribution function, we have with the notation (\ref{eq:indexnotation}) in the absence of a condensate:
\begin{equation}
 f({\mbf p}) + \frac{1}{2} =  \int_0^\infty \frac{d p^0}{2\pi}\, F_{aa}(p^0,{\mbf p}). \quad
\label{eq:nonfrel}
\end{equation} 
Comparison with the relativistic case shows that a difference in the scaling behavior is caused by the additional factor of $\sim p^0$ in the integrand of (\ref{eq:frel}). Therefore, we find that 
\begin{eqnarray}
&& f({\mbf p})  \, = \, s^{\kappa_{\mathrm s} + 2 - z} f(s{\mbf p}) \nonumber\\ 
&& \sim
\left\{ \begin{array}{ll}
|{\mbf p}|^{-(d + 2 - \eta)} & \mbox{\it nonrel.\ particle cascade}\\[0.1cm]
|{\mbf p}|^{-(d + 2 + z - \eta)} & \mbox{\it nonrel.\ energy cascade}
\end{array} \right.
\, \label{eq:nonrelfscal}
\end{eqnarray}
scales with one ``$z$'' difference than the relativistic solution (\ref{eq:relfscal}). This can have important consequences, such as the fact that the scaling exponent for the nonrelativistic particle cascade is now independent of the dynamic exponent $z$ describing the dispersion $\omega_{\mbf p} \sim |{\mbf p}|^z$. For $d=3$ and $\eta = 0$ one has the scaling $\sim 1/|{\mbf p}|^5$. This result from stationary turbulence is not far from the approximate power law $\sim 1/|{\mbf p}|^{\kappa_>}$ described in section~\ref{sec:outline} for the case of a self-similar evolution.

\section{Transport equations from quantum field theory\label{app:kadanoff}}

In this section, we discuss the most relevant approximations involved in the derivation of the kinetic equations used in the main text. We refer to Refs.~\cite{Berges:2010ez,Scheppach:2009wu,Aarts:2002dj,Berges:2005md,Gasenzer:2009ultra} for further details.

\subsection{Nonrelativistic transport equations}

We consider the time evolution of the statistical and spectral propagators $F$ and $\rho \equiv i \tilde{\rho}$ defined in (\ref{eq:stat_fct}) and (\ref{eq:spectral_fct}), which obey
\begin{align}
	[i\sigma_{ac}^3\partial_{x^0}-M_{ac}(x)]F_{cb}(x,y)=&\int_{t_0}^{x^0}dz\,\Sigma_{ac}^\rho(x,z)F_{cb}(z,y)\nonumber\\
	&-\int_{t_0}^{y^0}dz\,\Sigma_{ac}^F(x,z)\rho_{cb}(z,y),
\label{eq:evol_eq_F_nonrel}\\
	[i\sigma_{ac}^3\partial_{x^0}-M_{ac}(x)]\rho_{cb}(x,y)=&\int_{y^0}^{x^0}dz\,\Sigma_{ac}^\rho(x,z)\rho_{cb}(z,y),
\label{eq:evol_eq_rho_nonrel}
\end{align}
where
\begin{align}
	M_{ab}(x)=&\, \delta_{ab}\left[ -\frac{\nabla^2}{2m} + \frac{g}{2}\,\Big( F_{cc}(x,x) + \psi_c(x)\psi_c^*(x) \Big) \right] \nonumber\\
	&+ g\,\Big( F_{ab}(x,x) + \psi_a(x)\psi_b^*(x) \Big)
\end{align}
and $\psi_a(x)=\langle\hat{\psi}_a(x)\rangle$ is the macroscopic field, computed from complex nonrelativistic quantum fields $\hat{\psi}_1=\hat{\psi}$ and $\hat{\psi}_2=\hat{\psi}^\dagger$. 

Without approximating the self-energies, the above equations would be exact for Gaussian initial conditions. The approximate expressions for the statistical and spectral parts of the self-energy $\Sigma^\rho$ and $\Sigma^F$ used in this paper are given in appendix~\ref{app:2pi}. An equation for the macroscopic field completes the set of equations of motion in the case of a nonzero expectation value, which will not be used in the following. It can be found in reference~\cite{Scheppach:2009wu}.

The derivation of the transport equation for $F$ (for $\rho$) follows standard procedures. One starts by switching the variables $x$ and $y$ in (\ref{eq:evol_eq_F_nonrel}) (in (\ref{eq:evol_eq_rho_nonrel})) and subtracting the new equation from the original one. We change to Wigner coordinates $X^\mu\equiv(x^\mu+y^\mu)/2$ and $s^\mu\equiv x^\mu-y^\mu$ and Fourier transform with respect to the relative coordinate according to
\begin{align}
	g(X,p)=&\, \int_{-\infty}^{\infty}ds^\mu\,e^{ip_\mu s^\mu}\,g\left(X+\frac{s}{2},X-\frac{s}{2}\right).
\end{align}
Then one performs a gradient expansion, i.e.~an expansion in orders of $\partial_{X^\mu}$ and $\partial_{p_\mu}$, keeping only the first order in this expansion. In particular, this implies that
\begin{align}
	\int d^{d+1}s_{xy}&\, e^{ip_\mu s_{xy}^{\mu}}\,\int d^{d+1}z\,f(x,z)\,g(z,y) \nonumber\\
	&\, \approx\, f(X,p)\,g(X,p)
\label{eq:grad_exp_conv}
\end{align}
for any two functions $f(x,y)$ and $g(x,y)$. Writing $X^0\equiv t$ and dropping the $\mbf{X}$-dependence due to spatial homogeneity, one arrives at the transport equations to leading order gradient expansion
\begin{align}
	\frac{\partial F_{ab}}{\partial t}(t,p)=&\, -i\sigma_{ac}^3\left[ F_{cd}(t,p)\Sigma_{db}^\rho(t,p) - \rho_{cd}(t,p)\Sigma_{db}^F(t,p) \right],
\label{eq:transport_F_nonrel}\\
	\frac{\partial \rho_{ab}}{\partial t}(t,p)=&\, 0.
\label{eq:transport_rho_nonrel}
\end{align}

\subsection{Relativistic transport equations}

To derive the transport equations for a relativistic scalar $n$-component theory, we start with the 2PI evolution equations for the statistical and spectral propagators
\begin{align}
[\delta_{ac}\,\Box_x+M_{ac}^2(x)]F_{cb}(x,y)=&-\int_{t_0}^{x^0}dz\,\Sigma_{ac}^\rho(x,z)F_{cb}(z,y)\nonumber\\
&+\int_{t_0}^{y^0}dz\,\Sigma_{ac}^F(x,z)\rho_{cb}(z,y)
\label{eq:evol_eq_F},\\
[\delta_{ac}\,\Box_x+M_{ac}^2(x)]\rho_{cb}(x,y)=&-\int_{y^0}^{x^0}dz\,\Sigma_{ac}^\rho(x,z)\rho_{cb}(z,y),
\label{eq:evol_eq_rho}
\end{align}
where $\phi_a(x)=\langle \hat{\varphi}_a(x) \rangle$ and we defined the mass function 
\begin{align}
	M_{ab}^2(x)=&\, \delta_{ab}\left[m^2+\frac{\lambda}{6N}\Big( F_{cc}(x,x) + \phi_c(x)\phi_c(x) \Big) \right] \nonumber\\
	&+ \frac{\lambda}{3N}\Big( F_{ab}(x,x) + \phi_a(x)\phi_b(x) \Big)\,.
\label{eq:mass_function_rel}
\end{align}
The relativistic statistical and  spectral functions $F$ and $\rho$ are defined in analogy to equations~(\ref{eq:stat_fct}) and (\ref{eq:spectral_fct}) by substituting the operators $\hat{\psi}_a$ by scalar field operators $\hat{\varphi}_a$. Performing a gradient expansion to lowest order as explained in the previous subsection, one obtains the transport equations
\begin{align}
2p^0\frac{\partial F_{ab}}{\partial t}(t,p)&=-i\left[\Sigma_{ac}^\rho(t,p)F_{cb}(t,p)-\Sigma_{ac}^F(t,p)\rho_{cb}(t,p)\right],
\label{eq:transport_F}\\
2p^0\frac{\partial \rho_{ab}}{\partial t}(t,p)&=0\,.
\label{eq:transport_rho}
\end{align}
We note that the spectral function $\rho$ is time independent for both relativistic and nonrelativistic theories at this order of the gradient expansion.

\subsection{2PI $1/N$ expansion to NLO\label{app:2pi}}

We consider here the 2PI $1/N$ expansion to NLO, where we consider the case of a vanishing macroscopic field, i.e. $\phi_a = 0$~\cite{Berges:2001fi}. The corresponding equations for a nonzero field expectation value can be found in~\cite{Aarts:2002dj}. 
See reference~\cite{Berges:2004yj} for an introductory presentation.

At NLO the self-energies are given by
\begin{align}
	\Sigma^{F}_{ab}(x)=&\, B\left( I^F(x,y)F_{ab}(x,y) - \frac{1}{4}I^\rho(x,y)\rho_{ab}(x,y) \right)\label{eq:selfF_nlo},\\
	\Sigma^{\rho}_{ab}(x)=&\, B\bigg( I^\rho(x,y)F_{ab}(x,y) +I^F(x,y)\rho_{ab}(x,y) \bigg),\label{eq:selfrho_nlo}
\end{align}
where $B_{\text{nr}}=-2g$ and $B_{\text{rel}}=-\lambda/3N$ are the nonrelativistic and relativistic values for the prefactor $B$,
\begin{align}
	I^F(x,y)=&\, \Pi^F(x,y) - \int_{t_0}^{x^0}dz\,I^\rho(x,z)\Pi^F(z,y) \nonumber\\
	&\, + \int_{t_0}^{y^0}dz\,I^F(x,z)\Pi^\rho(z,y),
\label{eq:if_implicit}\\
	I^\rho(x,y)=&\, \Pi^\rho(x,y) - \int_{y^0}^{x^0}dz\,I^\rho(x,z)\Pi^\rho(z,y),\label{eq:irho_implicit}
\end{align}
and
\begin{align}
	\Pi^F(x,y)=&\, -\frac{B}{2}\text{Tr}\left[ F(x,y)F(y,x) + \frac{1}{4}\rho(x,y)\rho(y,x) \right],\\
	\Pi^\rho(x,y)=&\, -B\, \text{Tr}\left[ \rho(x,y)F(y,x)\right]
\label{eq:pi_rho_def}.
\end{align}
Here $F$ and $\rho$ are to be understood as matrices. Next, the initial time is sent to the remote past $t_0 \rightarrow -\infty$, and we introduce retarded and advanced quantities as $I^R(x,y)=\Theta(x^0-y^0)I^\rho(x,y)$, $I^A=-\Theta(y^0-x^0)I^\rho(x,y)$ and similar for $\Pi^R$ and $\Pi^A$. Then (\ref{eq:irho_implicit}) is Fourier transformed with respect to the relative Wigner coordinate to obtain at lowest order in the gradient expansion
\begin{align}
	I^R(X,p)=&\,\frac{\Pi^R(X,p)}{1+\Pi^R(X,p)}.
\end{align}
Using $I^\rho=I^R-I^A$ and proceeding in the same way with (\ref{eq:if_implicit}), one obtains
\begin{align}
	I^F(X,p)=&\, v_{\text{eff}}(X,p)\, \Pi^F(X,p) \\
	I^\rho(X,p)=&\, v_{\text{eff}}(X,p)\, \Pi^\rho(X,p)
\end{align}
with
\begin{equation}
	v_{\text{eff}}(X,p)=\frac{1}{|1+\Pi^R(X,p)|^2}
\label{eq:effective_coupling}
\end{equation}
and
\begin{align}
	\Pi^R(X,p)=&\, -B \int_q F_{ab}(X,q-p)\, G^R_{ba}(q).
\label{eq:pir_nonrel}
\end{align}
Inserting these expressions into the transport equation (\ref{eq:transport_F_nonrel}) and (\ref{eq:transport_F}), one obtains the evolution equations (\ref{eq:kineq_offshell_nonrel}) and (\ref{eq:kineq_offshell_rel}), respectively.

The effective coupling can be further simplified in the approximation with (\ref{eq:offshell_partnumb_def}), the identity
\begin{align}
	G^R_{ba}(p)=&\, \lim_{\epsilon\rightarrow0} \int \frac{d\omega}{2\pi} \,\frac{-i\rho_{ba}(\omega,\mbf p)}{\omega-p^0-i\epsilon}\,,
\label{eq:gr_rho}
\end{align}
and the on-shell spectral functions (\ref{eq:free_rho_nonrel_high}) and (\ref{eq:freerho}) for the nonrelativistic and relativistic theories, respectively. In this way, one arrives at the quasi-particle expressions (\ref{eq:pir_onshell_nonrel1}) and (\ref{eq:pir_onshell_rel}).

\section{Extracting scaling exponents for a self-similar evolution \label{app:error}}

In sections~\ref{sec:nonrel} and \ref{sec:rel} we observe that the nonrelativistic and relativistic $N=2$ and $N=4$ theories evolve in a self-similar way, characterized by (\ref{eq:selfsim}). We extract the pair of scaling exponents $(\alpha,\beta)$ and estimate their uncertainties from this self-similar evolution. For our analysis, we use a similar least-squares method as introduced in reference~{\cite{Berges:2013fga}}. In the following, we describe the method used in the present paper.

At first, we quantify the deviation from the self-similar evolution of the distribution function. Using the notation $\bar{p} \equiv |{\mbf p}|$, for each pair $(\alpha,\beta)$ the spectra at several times are rescaled according to 
\begin{eqnarray}
  f_{\text{resc}}(t, \bar{p}) = (t/t_{\text{ref}})^{-\alpha} f(t,(t/t_{\text{ref}})^{-\beta} \bar{p})\,.
\end{eqnarray}
We use the distribution at the earliest time $t_{ref}$ as a reference and the distributions at the $N_{\text{com}}$ later times for comparison. A perfectly self-similar evolution (\ref{eq:selfsim}) implies $\Delta f(t,\bar{p}) \equiv f_{\text{resc}}(t, \bar{p}) - f(t_{\text{ref}}, \bar{p}) = 0$ such that the rescaled distribution function becomes time independent. 

However, this equality is in general only true for the correct set of scaling exponents and is in practice violated due to statistical uncertainties of the data and systematic deviations from the perfect scaling behavior. Minimizing the deviations yields the best fit for the scaling exponents, and the study of the distribution of the deviations gives an estimate for the uncertainties of the exponents. We quantify the deviations by 
\begin{align}
  \chi^2(\alpha,\beta) = \frac{1}{N_{\text{com}}} \sum^{N_{\text{com}}}_{k=1} \frac{\int d(\log(\bar{p})) \left(\Delta f(t_k,\bar{p}) / f(t_{\text{ref}}, \bar{p})\right)^2}{\int d(\log(\bar{p}))},
\label{eq:chisqr}
\end{align}
where we sum over all relative deviations for each comparison spectrum. We use integration over $d(\log(\bar{p}))$ to increase the sensitivity at low momenta. The upper limit of integration is given by the highest momentum included in the inverse particle cascade, i.e.~the last point of the approximate $1/\bar{p}^{\kappa_>}$ power law of the reference distribution function. Since the spectra are binned in momentum space, the integrals translate to sums over momenta $\int d(\log(\bar{p})) \rightarrow \sum_{i=1}^{n_k-1} \log(\bar{p}_{i+1}/\bar{p}_i)$, where $\bar{p}_{i+1} > \bar{p}_i$ are the discrete momenta of each testing spectrum $k$ with possibly a different number of bins $n_k$. To compute the difference $\Delta f(t_k,\bar{p}_i)$,  we linearly interpolate momenta of the reference spectrum to coincide with the discrete momenta of the rescaled spectra.

The deviation $\chi^2(\alpha,\beta)$ is minimal for the best fit of the scaling exponents $\bar{\alpha}$ and $\bar{\beta}$. In analogy to reference~{\cite{Berges:2013fga}}, we define the likelihood function of a given set of exponents $(\alpha,\beta)$ as
\begin{align}
  W(\alpha,\beta) = \frac{1}{\mathcal{N}} \exp\left[ - \frac{\chi^2(\alpha,\beta)}{ 2\;\chi^2(\bar{\alpha},\bar{\beta})} \right],
\end{align}
where $\mathcal{N}$ is a normalization constant such that the integral $\int d\alpha\, d\beta\, W = 1$. Integrating $W(\alpha,\beta)$ over one of the exponents provides a marginal likelihood function for the other one, e.g.~$W(\alpha) = \int d\beta\, W(\alpha,\beta)$. Approximating the marginal likelihood functions with Gaussian distributions, we obtain an estimate for the standard deviations $\sigma_{\alpha}$ and $\sigma_{\beta}$, while the means are still given by $\bar{\alpha}$ and $\bar{\beta}$. Our thus measured scaling exponents are finally written in the form 
\begin{align}
  \alpha = \bar{\alpha} \pm \sigma_{\alpha}\;,\quad \beta = \bar{\beta} \pm \sigma_{\beta}\;.
\end{align}

\bibliography{mybibliography_alphabeta}

\end{document}